\documentclass[journal]{IEEEtran}
\usepackage{amsbsy}
\usepackage{bm}
\usepackage{booktabs}
\usepackage{amssymb}
\usepackage{amsmath}
\usepackage{amsthm}
\usepackage{comment}
\usepackage{hyperref}
\usepackage{tikz}
\usepackage{algorithm}
\usepackage{algorithmic}
\usepackage{url}
\usetikzlibrary{graphs, positioning, quotes, shapes.geometric}
\usepackage[margin=1in]{geometry}

\newtheorem{thm}{Theorem}
\newtheorem{lem}[thm]{Lemma}

\newcommand{\br}{{\mathbb R}}

\newcommand{\diag}{{\rm diag}}

\newcommand{\otherwise}{\mbox{\rm otherwise~}}
\newcommand{\prf}{\par{\bf Proof. }}
\newcommand{\bbox}{\hfill\rule{2mm}{3mm}}

\newcommand{\spann}{{\rm span}}
\newcommand{\st}{{\rm s.~t.~}}

\newcommand{\mcg}{\mathcal{G}}
\newcommand{\mcv}{\mathcal{V}}
\newcommand{\mce}{\mathcal{E}}

\newcommand{\mcx}{\mathcal{X}}

\newcommand{\bfu}{\mathbf{u}}

\newcommand{\bfh}{\mathbf{h}}
\newcommand{\bfg}{\mathbf{g}}

\newcommand{\bff}{\mathbf{f}}
\newcommand{\bfx}{\mathbf{x}}
\newcommand{\bfy}{\mathbf{y}}
\newcommand{\bfA}{\mathbf{A}}
\newcommand{\bfD}{\mathbf{D}}
\newcommand{\bfF}{\mathbf{F}}
\newcommand{\bfB}{\mathbf{B}}
\newcommand{\bfL}{\mathbf{L}}
\newcommand{\bfU}{\mathbf{U}}
\newcommand{\bfV}{\mathbf{V}}
\newcommand{\bfI}{\mathbf{I}}
\newcommand{\bfJ}{\mathbf{J}}

\newcommand{\bfQ}{\mathbf{Q}}
\newcommand{\bfR}{\mathbf{R}}
\newcommand{\bfH}{\mathbf{H}}
\newcommand{\bfW}{\mathbf{W}}

\def\argmax{\mathop{\mathgroup\symoperators argmax}}
\def\st{\mathrm{s. t.~}}

\title{Perfect Reconstruction Two-Channel Filter Banks on Arbitrary 
	Graphs Based on an Optimization Model}

\begin{document}
\twocolumn[
\begin{@twocolumnfalse}
		\section*{Perfect Reconstruction Two-Channel Filter Banks on Arbitrary Graphs Based on an Optimization Model\\[25pt]}
		
		``This work has been submitted to the IEEE for possible publication. Copyright may be transferred without notice, after which this version may no longer be accessible.''
\end{@twocolumnfalse}
]

\newpage
	
\author{Junxia You and Lihua Yang
	
	\thanks{This is supported by National Natural Science Foundation of China 
		(Nos. 12171488, 11771458) and Guangdong Province Key Laboratory of Computational Science 
		at the Sun Yat-sen University (2020B1212060032). (*Corresponding author: Lihua Yang).}
	\thanks{Junxia You is with School of Mathematics, Sun Yat-sen University, Guangzhou, China (e-mail: youjx3@mail2.sysu.edu.cn).} 
	\thanks{Lihua Yang is with School of Mathematics, Sun Yat-sen University, Guangzhou, China and 
		Guangdong Province Key Laboratory of Computational Science (e-mail: mcsylh@mail.sysu.edu.cn).}
}

\markboth{IEEE TRANSACTIONS ON SIGNAL PROCESSING, 2022}%
{Junxia You, Lihua Yang \MakeLowercase{\textit{et al.}}: Perfect Reconstruction Two-Channel Filter Banks on Arbitrary Graphs Based on an Optimization Model}
\maketitle
\begin{abstract}
In this paper, we propose the construction of critically sampled perfect reconstruction two-channel filterbanks on arbitrary undirected graphs. 
Inspired by the design of graphQMF proposed in the literature, 
we propose a general ``spectral folding property'' similar to that of bipartite graphs and provide sufficient conditions for constructing perfect reconstruction filterbanks based on a general graph Fourier basis, which is not the eigenvectors of the Laplacian matrix. To obtain the desired graph Fourier basis, we need to solve a series of quadratic equality constrained quadratic optimization problems (QECQPs) which are known to be non-convex and difficult to solve. We develop an algorithm to obtain the global optimal solution within a pre-specified tolerance. Multi-resolution analysis on real-world data and synthetic data are performed to validate the effectiveness of the proposed filterbanks.
\end{abstract}

\begin{IEEEkeywords}
	Graph signal processing, graph wavelet filterbank, graph Fourier basis, optimization, QECQP
\end{IEEEkeywords}

\section{Introduction}
\IEEEPARstart{G}{raphs} provide a natural representation for data domain in many applications, such as the social networks, web information analysis, sensor networks and machine learning. The collections of samples on these graphs are termed as \textit{graph signals}. For example, a social network can be modeled as a graph by viewing the individual accounts as vertices, the relationships between them as edges, and a certain attribute of all the accounts as a graph signal. Similarly, in a sensor network, the sensors, the distance between each two of them and the recorded data constitute a graph and a graph signal residing on it. In such applications, the underlying graphs typically have large sizes (the number of vertices), which poses challenges for signal transmission, analysis and storage, etc.
In classical signal processing, tools such as the wavelet filterbanks are employed to perform multi-resolution analysis on signals to achieve sufficient order reduction.
 For example, a two-channel filterbank decomposes a signal into a coarser approximation of the original signal and a detail signal that describes the complementary information of the original signal. However, due to the irregularity of graph structures, some traditional notions such as translation and dilation are difficult to establish on the graph setting. People are still actively seeking ways to develop wavelet transforms on graphs.

	In \cite{crovella2003graph}, Crovella and Kolaczyk constructed a series of simple functions on each neighbourhood of every vertex so that they are compactly supported and have zero integral over the entire vertex set, and the $l_2$-norms of these functions over the vertex set are all $1$.  They refer to these functions as graph wavelet functions. 
Coifman and Maggioni proposed the concept of diffusion wavelets and use diffusion as a smoothing and scaling tool to enable coarse graining and multiscale analysis in \cite{coifman2006diffusion}.
Gavish et al. \cite{gavish2010multiscale} first constructed multiscale wavelet-like orthonormal bases on hierarchical trees. They proved that function smoothness with respect to a metric induced by the tree is equivalent to approximate sparsity. Hammond et al. \cite{hammond2011wavelets} constructed wavelet transforms in the graph domain based on the spectral graph theory, and they presented a fast Chebyshev polynomial approximation algorithm to improve efficiency. In follow-up work, they also built an almost tight wavelet frame based on the polynomial filters \cite{tay2017almost}. All of these transforms are not critically sampled, i.e. the output of the transform oversamples the signal, which leads to the waste of space for storing the redundant information.
Narang and Ortega  overcame this drawback by designing filters called graphQMFs and critically sampled perfect reconstruction filter banks on bipartite graphs based on the  
spectral folding property. For arbitrary graphs, they proposed an algorithm that can decompose any graph into a collection of bipartite graphs, thereby extending the design to arbitrary graphs \cite{narang2012perfect}. In their follow-up work, they also constructed compactly supported biorthogonal wavelet filter banks \cite{Narang2013Compact}.  Based on the work of Narang and Ortega, Tanaka and Sakiyama \cite{tanaka2014Mchannel} proposed a scheme that uses an oversampled graph Laplacian matrix and proposed M-channel oversampled perfect reconstruction filterbanks on bipartite graphs. These oversampled filterbanks allow one to design filters more flexibly.
There are other critically sampled perfect reconstruction filterbanks, such as the spline-like wavelet filterbanks proposed by Ekambaram et al. \cite{Ekambaram2015Spline}, which was later extended to higher-order and exponential spline filters by Kotzagiannidis and Dragotti on circulant graphs \cite{2016Splines}.
\par	
Inspired by \cite{narang2012perfect}, we propose a method to construct  perfect reconstruction two-channel filterbanks on arbitrary graphs, without decomposing the given graph into a collection of bipartite subgraphs. 
Since the key to the design in \cite{narang2012perfect} is the spectral folding property of the Laplacian eigenvectors of bipartite graphs, which no longer holds for the general graphs,	we propose an algorithm to produce a new graph Fourier basis instead of using the Laplacian eigenvectors. Based on this Fourier basis, we establish a ``spectral folding property'' for general graphs similar to that for bipartite graphs and provide sufficient conditions for filters to satisfy the perfect reconstruction equation. It is worth mentioning that, in order to obtain the newly designed graph Fourier basis, we need to solve a series of quadratic equality constrained quadratic optimization problems (QECQPs). A QECQP is konwn to be non-convex and thus is difficult to find the global optimal solution. But in the context of our problem, we develop an algorithm to obtain the global optimal solution within a pre-specified tolerance. 
	
The rest of this paper is organized as follows:
In Section \ref{sec:background}, we intrduce some basic concepts such as the graph Fourier transform, filters, downsampling and upsampling and the two-channel filter banks. The related work \cite{narang2012perfect} is also introduced to motivate our work. In Section \ref{sec:proposed}, we present the 
``spectral folding property" in a general way and provide sufficient conditions for constructing perfect reconstruction filterbanks based on it. We introduce the theory and corresponding algorithms to compute the new graph Fourier basis as decirbed above. Then the construction of filters is based on this Fourier basis.  
	Experiments are also conducted to evaluate the performance of the proposed filterbanks and to compare with the related work in Section \ref{sec:expe}.

\section{Preliminaries}
\label{sec:background}

\subsection{Notations}
In this paper, vectors and matrices are represented by bold lowercase and bold uppercase letters, respectively. The $i$-th component of a vector $\bfx$ is denoted by $x_i$ or $\bfx(i)$. The $(i,j)$-th element of a matrix $\bfW$ is denoted by $w_{ij}$ or $\bfW(i,j)$. We write $\bfW\succeq0$ if the matrix $\bfW$ is positive semi-definite. Let $\mathbf{1}_N, \mathbf{0}_N$ and $\bfI_N$ represent the all-ones vector, the null vector and the identity matrix of order $N$, respectively.
A superscript $^\top$ denotes the transpose of a matrix or a vector. Function $\diag(\cdot)$ maps a vector to a diagonal matrix, or a matrix to its diagonal. $\|\bfx\|_2$ is the 2-norm of vector $\bfx$.
\par
We consider a connected, undirected and weighted graph $\mcg:=\{\mcv,\mce,\bfW\}$ where 
$\mcv$ and $\mce$ represent the sets of vertices and the set of edges of $\mcg$ respectively. 
We also assume that $\mcg$ has no self-loops and multi-edges.
Let $\mcv=\{v_1,...,v_N\}$ and 
$\bfW\in\br^{N\times N}$ be the corresponding adjacency matrix. Then  $w_{ij}>0$ if the vertices $v_i$ and $v_j$ are connected, and $w_{ij}=0$ otherwise. 
A graph signal $f:\mcv\rightarrow \br$ is a function defined on the vertices of the graph,
which can also be written as a vector $\bff\in\br^N$ with $i$-th component being $f(v_i)$.
In this paper, we will not distinguish the difference between $f$ and $\bff$ if no confusion arises. We adopt the following Dirichlet form to measure the oscillation of a graph signal $\bff$ with respect to (w.r.t) $\mcg$ \cite{shuman2013emerging}:
\[
S_2(\bff):=\sum_{i=1}^N\sum_{j=1}^{N}w_{ij}|f_i-f_j|^2.
\]
It is easy to see that the larger the value of $S_2(\bff)$, the stronger the signal oscillates on $\mcg$.

\subsection{Graph Fourier Transform and Filters}
\label{sec:GFT-filter}
The combinatorial Laplacian matrix of $\mcg$ is defined as $\bfL:=\bfD-\bfW$ where $\bfD:=\diag(d_1,...,d_N)$ with $d_i:=\sum_{j=1}^{N}w_{ij}$ is called the degree matrix of $\mcg$.
 As $\bfL$ is real symmetric, its Shur decomposition is given as $\bfL=\bfU\Lambda\bfU^\top$, where $\Lambda$ is a diagonal matrix: $\Lambda:=\diag(\lambda_1,...,\lambda_N)$ and 
 $0=\lambda_1<\lambda_2\leq\cdots\leq\lambda_N$ are all the eigenvalues of $\bfL$. 
 We call  the set $\sigma(\bfL):=\{\lambda_i\}_{i=1}^N$ the spectrum of $\bfL$. It is easy to see that
 the $i$-th column of $\bfU$, denoted by $\bfu_i$, is an eigenvector associated with eigenvalue $\lambda_i$ \cite{chung1997spectral}. 
	
	The eigenvectors $\{\bfu_i\}_{i=1}^N$ are often used as the graph Fourier basis, and the subsequent graph Fourier transform (GFT) of signal $\bff$ is defined as $\hat{\bff}:=\bfU^\top\bff$ \cite{shuman2013emerging}, 
	We notice that  the Dirichlet form of $\bff$ can also be written as
	\[
	S_2(\bff)=\bff^\top\bfL\bff.
	\]
	Since $S_2(\bfu_i)=\bfu_i^\top\bfL\bfu_i=\lambda_i$, we have $S_2(\bfu_1)\leq\cdots\leq S_2(\bfu_N)$. This implies that the oscillation of $\bfu_i$ intensifies as $i$ increases.
	In fact, the eigenvectors of $\bfL$ can be obtained by solving a series of optimization problems:
	\begin{equation}\label{opt:eigenofL}
		\begin{cases}
			\min\limits_{\substack{\bfx\perp \mathcal{U}_{k-1}\\\bfx\in\br^N,\|\bfx\|_2=1}}S_2(\bfx)
		\end{cases},~~k=1,...,N,
	\end{equation}
	where $\mathcal{U}_0=\{\bf0\}$ and $\mathcal{U}_k:=\{\bfu_1,...,\bfu_k\}$ when $k>0$, and $\bfu_k$ is the optimal solution to the $k$-th problem.
\par	
There are some other definitions of graph Fourier basis. In \cite{narang2012perfect,Narang2013Compact}, it is defined as the eigenvectors of the symmetric normalized Laplacian matrix $\mathcal{L}:=\bfD^{-1/2}\bfL\bfD^{-1/2}$. In \cite{sandryhaila2013discrete,sandryhaila2014big}, authors use the eigenvectors of the adjacency matrix $\bfW$ to form the  Fourier basis. In our previous work \cite{yang2021graph}, we propose a method to compute the graph Fourier basis by minimizing the $\ell^1$ oscillation of signals 
and such basis is shown to have good sparsity. Generally, any orthonormal basis 
$\bfu_1,...,\bfu_N$ satisfying $S(\bfu_1)\le...\le S(\bfu_N)$ 
for some oscillation measurement $S(\bfx)$ w.r.t graph $\mcg$ can be a regarded as a Fourier basis for the space of graph signals. Therefore, we can formulate a series of optmization problems similar to \eqref{opt:eigenofL} to caculate a graph Fourier basis.

\par
Given a graph Fourier basis $\bfu_1,...,\bfu_N$, we can define the filtering operation.
It is the modulation of the GFT of a graph signal:
\[
\bff\stackrel{\mbox{\scriptsize FT}}{\longrightarrow}
\hat{\bff}\stackrel{\mbox{\scriptsize M}}{\longrightarrow}
\big[h_1\hat{\bff}_1, ..., h_N\hat{\bff}_N\big]^\top
\stackrel{\mbox{\scriptsize IFT}}{\longrightarrow}\bfF_h\bff,
\]
or equivalently $\bfF_h=\bfU\diag(\bfh)\bfU^\top$,
where $\bfU=[\bfu_1,...,\bfu_N]$ is called the Fourier basis matrix and FT, IFT, M are the abbreviations of ``Fourier transform'', ``Inverse Fourier Transform'' and ``Modulation''. 
The vector $\bfh:=(h_1,..., h_N)^{\top}$ used for frequency modulation is called the filter vector.

\subsection{Downsampling and Upsampling}
\label{sec:sampling}
A downsampling operation we consider is a map defined on the graph $\mcg=\{\mcv,\mce,\bfW\}$ that choose a subset $\mcv_L\subset \mcv$ such that the samples of graph signal $\bff$ indexed in $\mcv_L$ are retained while the rest are discarded. Mathematically, it can be written as the matrix $\bfA_L\in\br^{|\mcv_L|\times N}$ consisting of rows of $\bfI_N$ whose indices are in $\mcv_L$. The corresponding upsampling operation maps the downsampled graph signal back into $\br^N$ by inserting zeros on the indices in $\mcv_L^c:=\mcv\backslash\mcv_L$. So the upsampler can be written as the matrix $\bfB_L\in\br^{N\times|\mcv_L|}$ consisting of columns of $\bfI_N$ whose indices are in $\mcv_L^c$. Correspondingly we can define the downsampler $\bfA_H$
and upsampler $\bfB_H$ for the complementary set $\mcv_H:=\mcv_L^c$. It is easy to verify that
$\bfB_L=\bfA_L^\top$ and $\bfB_H=\bfA_H^\top$. 
\par
The pair of subsets $\{\mcv_L, \mcv_H\}$ satisfying $\mcv_L\cup\mcv_H=\mcv$ and 
$\mcv_L\cap\mcv_H=\emptyset$  is called a partition of $\mcv$. 
It is easy to see that the corresponding downsampling then upsampling (DU) operation can be expressed as 
\begin{equation}\label{eq:DUbyJ}
\bfB_L\bfA_L=\frac{1}{2}(\bfI_N+\bfJ),~~\bfB_H\bfA_H=\frac{1}{2}(\bfI_N-\bfJ),
\end{equation}
where $\bfJ:=\diag(J_{11},...,J_{NN})$ with 
\begin{align}\label{def:J}
J_{ii}=\begin{cases}
1,&v_i\in\mcv_L,\\
-1,&\text{otherwise},
\end{cases}
\end{align} 
is called the sampling matrix associated with the partition $\{\mcv_L, \mcv_H\}$. Obviously, the samplers or the sampling matrix is determined by the partition $\{\mcv_L, \mcv_H\}$.

\subsection{Two-Channel Filterbanks and Related Work}
\label{sec:TCFB}
\begin{figure}[hbtp]
	\begin{center}
		\tikzstyle{basic} = [rectangle, minimum width=0.8cm,minimum height=0.5cm,text centered,draw=black]
		\tikzstyle{du} = [rectangle, rounded corners, minimum width=0.8cm,minimum height=0.5cm,text centered,draw=black]
		\tikzstyle{arrow} = [thick,->,>=stealth]
		\begin{tikzpicture}[node distance=10pt]
			\node[rounded corners,draw=black] (x)  {$\bfx$};
			\node[basic,above right=15pt of x,yshift=-0.5cm]   (DL)  {$\bfF_{h_0}$};
			\node[du,right=15pt of DL]                (SL)  {$\bfA_L$};
			\node[du,right=15pt of SL]						(SLU){$\bfB_L$};
			\node[basic,right=15pt of SLU]						(RL){$\bfF_{g_0}$};
			\node[basic,below right=15pt of x,yshift=0.5cm]            (DH)  {$\bfF_{h_1}$};
			\node[du,right=15pt of DH]                        (SH)  {$\bfA_H$};
			\node[du,right=15pt of SH]						(SHU){$\bfB_H$};
			\node[basic,right=15pt of SHU]						(RH){$\bfF_{g_1}$};			
			\node[rounded corners, draw=black,right=160pt of x]  (y)     {$\bfy$};

			\draw [arrow] (x) |- (DL);
			\draw [arrow] (x) |- (DH);
			\draw [arrow] (DL) -- (SL);
			\draw [arrow] (DH) -- (SH);
			\draw [arrow] (SL) -- (SLU);
			\draw [arrow] (SH) -- (SHU);
			\draw [arrow] (SLU) -- (RL);
			\draw [arrow] (SHU) -- (RH);
			\draw [arrow] (RL) -| (y);
			\draw [arrow] (RH) -| (y);
		\end{tikzpicture}
	\end{center}
	\caption{A two-channel filter bank.}\label{FB}
	\label{fig:2-channel-banks}
\end{figure}
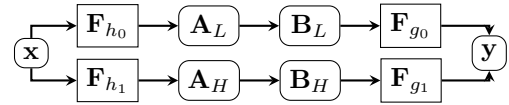

A two-channel filter bank consists of two lowpass 
filters $\bfF_{h_0}$ and $\bfF_{g_0}$, two highpass filters $\bfF_{h_1}$ and $\bfF_{g_1}$, two downsamplers $\bfA_L, \bfA_H$ and two upsamplers $\bfB_L, \bfB_H$. 
The filters $\bfF_{h_0}$ and $\bfF_{h_1}$ are called analysis filters, and the filters $\bfF_{g_0}$ 
and $\bfF_{g_1}$ are called synthesis filters.
With a two-channel filter bank, the input signal $\bfx$ is separated into two frequency bands, a low frequency band in the lowpass channel, and a high frequency band in the highpass channel. We use subscripts $_L$ and $_H$ to represent lowpass and highpass, respectively. The signal in each channel between the analysis and synthesis stages may be coded for 
transmission or storage, and compression may occur in which case information is lost. For perfect reconstruction, i.e., 
\begin{equation}\label{eq:pr-condition-0}
	\bfy=(\bfF_{g_0}\bfB_L\bfA_L\bfF_{h_0}+\bfF_{g_1}\bfB_H\bfA_H\bfF_{h_1})\bfx,
\end{equation}
we require the synthesis bank to connect directly to the analysis bank \cite{winkler2000orthogonal}. A flow chart is displayed in Figure  \ref{fig:2-channel-banks}. 

\par
By \eqref{eq:DUbyJ}, we can rewrite \eqref{eq:pr-condition-0} as
\begin{equation}\label{eq:PR}
\bfF_{g_0}(\bfI_N+\bfJ)\bfF_{h_0}+\bfF_{g_1}(\bfI_N-\bfJ)\bfF_{h_1}=2\bfI_N.
\end{equation}
To reduce the difficulty of solving the problem  we consider the following stronger condition:
\begin{equation}\label{eq:PR_2}
\bfF_{g_0}\bfF_{h_0}+\bfF_{g_1}\bfF_{h_1}=2\bfI_N,~~
\bfF_{g_0}\bfJ\bfF_{h_0}-\bfF_{g_1}\bfJ\bfF_{h_1}=\mathbf{0},
\end{equation}
which is called the perfect reconstruction condition hereafter. 
The goal of this paper is to design the sampling matrix $\bfJ$ and the filters $\bfF_{h_0},\bfF_{h_1},\bfF_{g_0},\bfF_{g_1}$ such that
\eqref{eq:PR_2} holds.
\par
The first equation of perfect reconstruction condition can be simplfied as
$\bfg_0\odot \bfh_0+\bfg_1\odot \bfh_1=2\mathbf{1}_N$, where $\odot$ stands for the Hadamard product of 
vectors. But the second equation cannot be simplified similarly since the matrices 
$\bfF_{g_0}, \bfF_{g_1}$ or $\bfF_{h_0},\bfF_{h_1}$ do not commute with the sampling 
matrix $\bfJ$ generally. For bipartite graphs with two parts of vertices $\mcv_1$ and $\mcv_2$, i.e., vertices in the same part are not connected, thus edges only exist between $\mcv_1$ and $\mcv_2$, let $\mathcal{L}_b$ be its symmetric normalized Laplacian matrix with eigendecomposition $\mathcal{L}_b=\bfV\Lambda_b\bfV^\top$, where $\Lambda_b:=\diag(\lambda_1^b,...,\lambda_N^b)$ and $0=\lambda_1^b<\lambda_2^b\leq\cdots\leq\lambda_N^b$ are the eigenvalues of $\mathcal{L}_b$. When the partition $\{\mcv_L, \mcv_H\}$ for downsapmpling and 
upsampling coincide with the two parts, Narang and Ortega  show the following equation \cite{narang2012perfect}:
\begin{equation}\label{eq:spectral-folding}
\bfJ\bfF_{h_0(\cdot)}=\bfF_{h_0(2-\cdot)}\bfJ,
\end{equation}
where the filter $\bfF_{h_0(\cdot)}$ is a spectral filter \cite{hammond2011wavelets} defined as 
\[
\bfF_{h_0(\cdot)}:=\bfV\diag(\bfh_0)\bfV^\top,
\]
where $\bfh_0=[h_0(\lambda_1^b),...,h_0(\lambda_N^b))]^\top$ and $h_0(\cdot)$ is a real funtion defined on the spectrum $\sigma(\mathcal{L}_b)$. The spectral folding property of bipartite graphs states that
\begin{equation}\label{eq:PJ_commu_bipar}
	\bfJ\bfV=\bfV\Phi_b,
\end{equation}
where $\Phi_b$ is the anti-diagonal identity matrix, which is also a permutation matrix. Then \eqref{eq:spectral-folding} is obtained as a direct inference.
In this case, we can simplify the perfect reconstructin condition using \eqref{eq:spectral-folding} as
\begin{align}\label{eq:prcon_2bipart}
	\begin{cases}
		g_0(\lambda)h_0(\lambda)+g_1(\lambda)h_1(\lambda)=2,\\
		g_0(\lambda)h_0(2-\lambda)g_1(\lambda)h_1(2-\lambda)=0,
	\end{cases}~~\lambda\in~[0,2].
\end{align}

\par	
However, the spectral folding property \eqref{eq:PJ_commu_bipar} generally dose not hold for an arbitrary graph, thus the perfect reconstruction condition cannot be simplified like \eqref{eq:prcon_2bipart}.  In \cite{narang2012perfect}, the authors formulate a bipartite subgraph decomposition algorithm to decompose a non-bipartite graph into a collection of bipartite subgraphs and the two-channel filterbanks are constructed on each subgraph. 
\par
It is challenging to construct perfect reconstruction filter bank on non-partite
graphs. There are two key designs that need to be done: (1) Construct the sampling matrix $\bfJ$ and then (2) solve the equations of perfect reconstruction condition \eqref{eq:PR_2} with the given sampling matrix.

\section{Proposed Method}
\label{sec:proposed}

As described in Section \ref{sec:TCFB}, In order to construct a perfect reconstruction 
two-channel filterbank on an arbitrary graph, we need to design a sampling matrix and filters that satisfy the perfect reconstruction condition \eqref{eq:PR_2}.

\subsection{Construction of the Sampling Set $\mcv_L$}
\label{sec:construction-samplingset}
We first talk about how to construct the sampling matrix $\bfJ$. 
 As discussed in Section \ref{sec:sampling}, it is sufficient to determine the partition $\{\mcv_L, \mcv_H\}$ of the vertex set $\mcv$.
 How to determine the partition has been studied in the literature, such as using the polarities of entries of $\bfu_N$ which is the largest eigenvector of $\bfL$, or doing $k$-means clustering on $\bfu_N$ \cite{Shuman2016Amultiscale}, or solving the max-cut problem \cite{Narang2010Local} to obtain $\mcv_L$ and hence $\mcv_H:=\mcv_L^c$.
 
In this paper, we consider the max-cut problem:
\[
\argmax_{\mcv_L}\{\mbox{cut}(\mcv_L,\mcv_L^c)\}
=\argmax_{\mcv_L}\Big\{\sum_{i\in\mcv_L}\sum_{j\in\mcv_L^c}w_{ij}\Big\},
\]
where $w_{ij}$ is the $(i,j)$-th element of the adjacency matrix $\bfW$. 
This is equivalent to 
\begin{align}
\begin{cases}
\max\limits_{\mathbf{d}_J}&\mathbf{d}_J^\top\bfL\mathbf{d}_J\\
\st&\mathbf{d}_J\in\{-1,1\}^N.
\end{cases}
\end{align}
	Thus, we consider such a greedy algorithm to determine $\mcv_L$. Start with an empty set $\mcv_L$. First select the vertex with the largest degree, denoted $v_0$, and add it to $\mcv_L$. Then repeatedly add the vertex $v_i$ that maximizes $S_i:=\sum_{x,y\in\mcv_L^i}\bfL_{\mcv_L^i,\mcv_L^i}(x,y)$ to $\mcv_L$, until $S_i$ starts to decrease, where $\mcv_L^i:=\mcv_L\cup\{v_i\}$ and $\bfL_{\mcv_L^i,\mcv_L^i}$ is the submatrix of $\bfL$ consisting of all the elements with row index and column index  in $\mcv_L^i$. After $\mcv_L$ is determined, the sapling matrix $\bfJ$ can be obtained according 
to \eqref{def:J}.  The pseudo-code is given in Algorithm \ref{alg:J}, where we denote $S:=\sum_{x,y\in\mcv_L}\bfL_{\mcv_L,\mcv_L}(x,y)$ and $S_i$ is as defined above.
\begin{algorithm}[htbp]\
		\renewcommand{\algorithmicrequire}{\textbf{Input:}}
		\renewcommand{\algorithmicensure}{\textbf{Output:}}
		\caption{Compute $\mcv_L$}
		\label{alg:J}
		\begin{algorithmic}[1]
			\REQUIRE Graph Laplacian matrix $\bfL$
			\STATE Initialization: Set $\mcv_L=\emptyset,\mcv^c_L=\mcv$
			\REPEAT 
			\STATE Compute $v:=\argmax_{v_i\in\mcv_L^c}~S_i$.
			\STATE Update $\mcv_L=\mcv_L\cup\{v\},\mcv^c_L=\mcv\backslash\mcv_L$
			\UNTIL $S_i<S$
			\STATE $\mcv_L=\mcv_L\backslash\{v\}$
			\ENSURE $\mcv_L$
		\end{algorithmic}
\end{algorithm}

\subsection{Constructing a Graph Fourier Basis Based on an Optimization Model}

\subsubsection{Optimization Model for Constructing a Graph Fourier Basis}

Suppose the sampling matrix $\bfJ$ is determined, we aim to find a graph Fourier basis matrix $\bfU$ and a permutation matrix $\Phi$ satisfying the general ``spectral folding property'': $\bfJ\bfU=\bfU\Phi$,
that is, $\bfJ\bfu_1,...,\bfJ\bfu_N$ is a permutation of $\bfu_1,...,\bfu_N$. Then, similar to the case of bipartite graphs, filters constructed by this $\bfU$ and any filter vectors can weakly commute with $\bfJ$, just like \eqref{eq:spectral-folding}, thus we can simplify the perfect reconstruction condition for an arbitrary graph.
For this purpose, we adopt the oscillation measurement $S_2(\bfx)=\bfx^\top\bfL\bfx$
to find a graph Fourier basis matrix $\bfU=[\bfu_1,...,\bfu_N]$ by 
solving the following optimization problems iteratively:
\begin{align}\label{opt:fourier}
	&\begin{cases}
		\min\limits_{\bfx\in\mathcal{X}_k}&S_2(\bfx)\\
		\st &\bfx^\top\bfx=1,\\
		&\bfx^\top\bfJ\bfx=0,
	\end{cases}
\end{align}
for $k=1, 2,..., K$,  where 
\[
\mathcal{X}_k:=\{\bfx\in\br^N|\bfU_{k-1}^\top\bfx=0\}=(\spann(\bfU_{k-1}))^\perp
\]
and $\bfU_0:=\mathbf{0}_N$ and
$\bfU_k:=[\bfu_1,...,\bfu_k, \bfJ\bfu_1,...\bfJ\bfu_k]$ with $\bfu_k$ being the optimal solution to the $k$-th problem \eqref{opt:fourier}, until the feasible set 
\begin{equation}\label{eq:feasibleset-k}
\mathcal{D}_k:=\{\bfx\in\mathcal{X}_k|\bfx^\top\bfx=1,~\bfx^\top\bfJ\bfx=0\}
\end{equation}
of Problem \eqref{opt:fourier} is empty, that is, $\mathcal{D}_K\neq\emptyset$
while $\mathcal{D}_{K+1}=\emptyset$.
\par
We claim that for $k=1,2,...,K+1$, the space $\mathcal{X}_k$ is $\bfJ$-invariant in the sense that $\bfJ\mathcal{X}_k=\mathcal{X}_k$ and $\dim\mathcal{X}_k=N-2(k-1)$. In fact, this is true for $k=1$ since $\mathcal{X}_1=\br^N$. 
Suppose that for $k\ge 1$, $\mathcal{X}_{k}$ is $\bfJ$-invariant and $\dim\mathcal{X}_{k}=N-2(k-1)$. 
Then, if $\mathcal{D}_{k}$ is non-empty, there must be an optimal solution to Problem \eqref{opt:fourier} since $\mathcal{D}_{k}$ is compact and the objective function is continuous. Let $\bfu_k$ be such a solution, then by definition $\mcx_{k+1}$ is the orthogonal complement of $\spann(\bfu_k,\bfJ\bfu_k)$ in the space $\mcx_k$, i.e., $\mcx_k=\spann(\bfu_k,\bfJ\bfu_k)\oplus\mcx_{k+1}$. Since $\bfJ$ is orthogonal and $\bfJ\mcx_k=\mcx_k$, we have $\spann(\bfu_k,\bfJ\bfu_k)\oplus\bfJ\mcx_{k+1}=\mcx_k$, thus $\bfJ\mcx_{k+1}=\mcx_{k+1}$ and $\dim\mathcal{X}_{k+1}=\dim\mathcal{X}_{k}-2$. By induction, the claim is true.
\par 
The following theorem provides the stopping criterion of iteration of Problem \eqref{opt:fourier}, i.e., when the feasible set $\mathcal{D}_k$ starts to be empty.
\begin{thm}\label{th:switchonA}
	For the $k$-th problem \eqref{opt:fourier} with $k=1,..., K+1$,
	 if $\dim\mathcal{X}_k>0$, 
	let $\bfA_k\in\br^{N\times\ell}$ be a matrix whose columns form an orthonormal 
	basis for $\mcx_k$, then the feasible set $\mathcal{D}_k$ is non-empty if and only if 
	the eigenvalues of $\bfA_k^\top\bfJ\bfA_k$ are $\{1,-1\}$.
\end{thm}
\prf
According to the above discussion we have $\bfJ\mathcal{X}_k=\mathcal{X}_k$. 
Since the columns of $\bfA_k$ form an orthonormal basis for $\mathcal{X}_k$, there must exist a matrix $\bfB$ such that $\bfJ\bfA_k=\bfA_k\bfB$ and hence $\bfB=\bfA_k^\top\bfJ\bfA_k$. Let $\bfA_k^\top\bfJ\bfA_k=\Xi_k\diag(\mu_1,...,\mu_\ell)\Xi_k^\top$ be
its eigen-decomposition, where $\Xi_k$ is an orthogonal matrix
and $\mu_1,...,\mu_\ell$ are all the eigenvalues of $\bfA_k^\top\bfJ\bfA_k$. Then
we have $\bfJ\bfA_k\Xi_k=\bfA_k\bfB\Xi_k=\bfA_k\Xi_k\diag(\mu_1,...,\mu_\ell)$. For $j=1,...,\ell$, it is 
easy to see that the $j$-th column of $\bfA_k\Xi_k$ is an eigenvector of $\bfJ$ 
associated with eigenvalue $\mu_j$. Thus $\mu_j=\pm 1$, which means that each eigenvalue 
of $\bfA_k^\top\bfJ\bfA_k$ must be $1$ or $-1$. 
\par
Since any $\bfx\in\mcx_k$ can be expressed as $\bfx=\bfA_k\xi$ for some 
$\xi\in\br^\ell$ with $\xi^\top\xi=\bfx^\top\bfx$, it is easy to see that
$\mathcal{D}_k$ is not empty if and only there exists $\xi\in\br^\ell$ such
that $\xi^\top\xi=1$ and $\xi^\top(\bfA_k^\top\bfJ\bfA_k)\xi=0$, that is,
the matrix $\bfA_k^\top\bfJ\bfA_k$  is neither positive 
definite nor negative definite, or equivalently, the eigenvalues of 
$\bfA_k^\top\bfJ\bfA_k$ are $\{1, -1\}$.
\par
\textbf{Remark:} it is easy to see that, if $\dim\mathcal{X}_k>0$ and all the eigenvalues of 
$\bfA_k^\top\bfJ\bfA_k$ are $1$s, then $\bfA_k^\top\bfJ\bfA_k=\bfI_\ell$. Similarly,
if $\dim\mathcal{X}_k>0$ and all the eigenvalues of 
$\bfA_k^\top\bfJ\bfA_k$ are $-1$s, then $\bfA_k^\top\bfJ\bfA_k=-\bfI_\ell$. 
%
%
\bbox
\par
Let us equivalently simplify the problems \eqref{opt:fourier} as follows when $\mathcal{D}_k$ is non-empty:
\begin{align}\label{opt:QECQP}
	\begin{cases}
		\min &S_2(\bfA_k\xi)\\
		\st&\xi^\top\xi=1,\\
		&\xi^\top(\bfA_k^\top\bfJ\bfA_k)\xi=0.
	\end{cases}
\end{align}
Then according to Theorem \ref{th:switchonA}, Problem \eqref{opt:QECQP} is feasible iff $\mathcal{D}_k$ is non-empty iff $\bfA_k^\top\bfJ\bfA_k\neq\pm\bfI_\ell$. Thus, we are going to solve Problem \eqref{opt:QECQP} iteratively until the following the stopping criterion:
\begin{align*}
	\dim\mathcal{X}_k=0,~~\mbox{or}~~
	\begin{cases}
		\dim\mathcal{X}_k>0,\\
		\bfA_k^\top\bfJ\bfA_k=\bfI_\ell~\mbox{or}~\bfA_k^\top\bfJ\bfA_k=-\bfI_\ell,
	\end{cases}
\end{align*}
 is satisfied
\par
(1) If $\dim\mathcal{X}_k=0$ then the columns of $\bfU_{k-1}$ already constitute an orthonormal
basis of $\br^N$.
\par
(2) If $\dim\mathcal{X}_k>0$ but
$\bfA_k^\top\bfJ\bfA_k=\bfI_\ell$ or $\bfA_k^\top\bfJ\bfA_k=-\bfI_\ell$,
then we find all the eigenvectors of $\bfA_k^\top\bfL\bfA_k$ and insert them into
$\bfU_{k-1}$ to form an orthonormal basis for $\br^N$.
\par
In summary, we can get a matrix $\bfU_{k-1}$ whose columns constitute an orthonormal 
basis for $\br^N$. This matrix is the desired graph Fourier basis matrix.
The pseudo-codes for computing all the Fourier basis vectors are shown in Algorithm \ref{alg:basis}. The function \textbf{SolveOpt} in the third step solves the optimization problem \eqref{opt:QECQP} in each iteration to find an optimal solution $\xi$, which will be discussed in detail later.
\begin{algorithm}[h]
	\renewcommand{\algorithmicrequire}{\textbf{Input:}}
	\renewcommand{\algorithmicensure}{\textbf{Output:}}
	\caption{Calculate the graph Fourier basis}
	\label{alg:basis}
	\begin{algorithmic}[1]
		\REQUIRE Sampling matrix $\bfJ\neq\pm\bfI_N$, graph Laplacian $\bfL$
		\STATE Initialization: Set $\mcx=\br^N,\bfA=\bfI_N,k=1$
		\REPEAT 
		\STATE Compute $\xi$=\textbf{SolveOpt}($\bfL,\bfJ,\bfA$)
		\STATE Set $\bfu_k=\bfA\xi,\bfu_{N-k+1}=\bfJ\bfu_k,\mcx=\mcx\ominus\spann\{\bfu_k,\bfu_{N-k+1}\}$
		\STATE Compute an orthonormal basis for $\mcx$ to form matrix $\bfA$
		\STATE $k=k+1$
		\UNTIL $\mcx$ is null or $\bfA^\top\bfJ\bfA$ is either positive or negative definite 
		\IF {$\mcx$ is not null}
		\STATE Let $\bfU_A$ be the matrix of eigenvectors of $\bfA^\top\bfL\bfA$
		\ELSE 
		\STATE Set $\bfU_A=[~]$
		\ENDIF
		\STATE Let $\bfU=[\bfu_1,\bfu_N,...,\bfu_k,\bfu_{N-k+1},\bfU_A]$ 
		\STATE Rearrange the columns of $\bfU$ in ascending order of $\{S_2(\bfu_i)\}_{i=1}^N$
		\ENSURE The Graph Fourier matrix $\bfU$
	\end{algorithmic}
\end{algorithm}

\subsubsection{Solving the Optimization Problem}
\label{sec:geQECQP}
\par
Problem \eqref{opt:QECQP} is a quadratic equality constrained quadratic optimization problem (QECQP). It is not an easy task to find the global optimal solution since it is not a convex 
problem. In this section, we shall develop an algorithm to compute the global optimal solution to Problem \eqref{opt:QECQP}, as described in Algorithm \ref{alg:solveopt} for \textbf{SolveOpt}.  
\par
Let us first consider the following more general QECQP, denoted by geQECQP:
\begin{align}\label{opt:gene_QECQP}
	\begin{cases}
		\min\limits_{\bfx\in\br^n}&\bfx^\top\bfQ\bfx\\
		s.t.&\bfx^\top\bfx=1,\\
		&\bfx^\top\bfR\bfx=1,
	\end{cases}
\end{align}
where $\bfQ,\bfR\in\br^{n\times n}$ and $\bfR$ is positive semi-definite. Denote the feasible set of the problem by
\[
\mathcal{D}:=\{\bfx\in\br^n|\bfx^\top\bfx=\bfx^\top\bfR\bfx=1\}.
\]
\par
The well-known Lagrangian function of the problem \eqref{opt:gene_QECQP} is given by
\begin{align*}
	\bfL(x,\mu_1,\mu_2)&=\bfx^\top\bfQ\bfx+\mu_1(\bfx^\top\bfx-1)
	+\mu_2(\bfx^\top\bfR\bfx-1)\\
	&=\bfx^\top\bfH(\mu_1,\mu_2)\bfx-\mu_1-\mu_2,
\end{align*}
where
\[
\bfH(\mu_1,\mu_2):=\bfQ+\mu_1\bfI_N+\mu_2\bfR.
\]
It is known that if $\bfx\in\mathcal{D}$ is a local optimal solution $\bfx$, then
the following necessary conditions hold:
\begin{itemize}
\item
First-order condition:
$\bfH(\mu_1,\mu_2)\bfx=\bf0$;
\item
Second-order condition:
$\bfy^\top\bfH(\mu_1,\mu_2)\bfy\geq0$ for any 
$\bfy\in T:=\{\bfy|\bfy^\top\bfx=0,\bfy^\top\bfR\bfx=0\}$. 
\end{itemize}
In \cite{bar1994global}, Bar-on and Grasse give a sufficient and necessary condition for a local optimal solution to be a global solution under the following assumptions:
\par
(a) $\bfR$ is positive semi-definite;
\par
(b) The eigenvalues of $\bfR$ spread across one;
\par
(c) The vectors $\bfR\bfx$ and $\bfx$ are linearly independent for each $\bfx\in\mathcal{D}$.
\begin{lem}\label{lemma}\cite{bar1994global}
Under the assumptions (a)--(c), 
a necessary and sufficient condition for a triplet $(\bfx;\mu_1,\mu_2)$ to yield the global optimal solution of geQECQP \eqref{opt:gene_QECQP} is that it satisfys the first- and second-order necessary conditions for optimality and that $\bfH(\mu_1,\mu_2)\succeq 0$.
\end{lem}
\par
\textbf{Remark: }
Condition (b) is equivalent to $\mathcal{D}\neq\emptyset$ \cite{bar1994global};
Condition (c) is equivalent to $\bfR\bfx\neq\bfx$ for any $\bfx\in\mathcal{D}$. In fact, for any 
$\bfx\in\mathcal{D}$, if $\bfR\bfx$ and $\bfx$ are linearly dependent 
then $c_1\bfx-c_2\bfR\bfx=\mathbf{0}_N$ for some $c_1, c_2\in\br$ satisfying $|c_1|+|c_2|\neq 0$. 
Pre-multiplying on both sides of the equation by $\bfx^\top$ yields $c_1=c_2\neq 0$, which implies that $\bfR\bfx=\bfx$. Conversely, $\bfR\bfx=\bfx$ implies the linear dependence of
$\bfR\bfx$ and $\bfx$. Hence Condition (c) is satisfied if $\bfR$ has no eigenvalues $1$.
\par
Lemma \ref{lemma} inspires us to find $\mu^*_1, \mu^*_2$ and $\bfx^*\in\mathcal{D}$ 
such that
\begin{equation}\label{eq:iffcondition}
	\bfH(\mu^*_1,\mu^*_2)\succeq 0~~~\mbox{and}~~~\bfH(\mu^*_1,\mu^*_2)\bfx^*=0.
\end{equation}
If such a triplet $(\bfx^*;\mu^*_1,\mu^*_2)$ is  found, then $\bfx^*$ is a global optimal solution of geQECQP. 
\par
To find such $\mu_1^*$ and $\mu_2^*$, we consider the dual problem
\begin{equation}\label{eq:dual-problem}
\eta^*:=\sup_{\mu_1,\mu_2\in\br}g(\mu_1,\mu_2),
\end{equation}
where
\begin{align*}
	g(\mu_1,\mu_2)&:=\inf_{\bfx\in\br^N}\bfL(x,\mu_1,\mu_2)\\
	&=\begin{cases}
		-\mu_1-\mu_2, &\bfH(\mu_1,\mu_2)\succeq 0,\\
		-\infty, & \otherwise,
	\end{cases}
\end{align*}
is the dual function of geQECQP. The well-known weak duality tells us that
$\eta^*\le\zeta^*$, where $\zeta^*:=\inf_{\bfx\in\mathcal{D}}(\bfx^\top\bfQ\bfx)$
is the optimal value of geQECQP \cite{boyd2004convex}. 
Obviously, \eqref{eq:dual-problem} is equivalent to
\begin{equation}\label{opt:dual}
	\eta^*:=\sup_{\bfH(\mu_1,\mu_2)\geq0}(-\mu_1-\mu_2).
\end{equation}	

In the rest of the section, we shall prove that
\begin{align*}
\begin{cases}
	\mu_1^*:=-\lambda_{\min}(\bfQ+\mu^*_2\bfR),\\
	\mu_2^*=\argmax_{\mu_2\in\br}\big[-\mu_2+\lambda_{\min}\big(\bfQ+\mu_2\bfR)\big]
\end{cases}
\end{align*}
is a solution of the dual problem \eqref{opt:dual} and
satisfies $\bfH(\mu^*_1,\mu^*_2)\succeq 0$.

\begin{thm}\label{lem:my1}
	Let $\bfx^*\in\br^N$. Then $\bfx^*$ be a solution of geQECQP \eqref{opt:gene_QECQP}
	if and only if there
	exist $\mu_2^*\in\br$ and $\mu_1^*=-\lambda_{\min}(\bfQ+\mu^*_2\bfR)$, such that 
	\begin{equation}\label{eq:lem:my1-1}
		-\mu_2^*+\lambda_{\min}\big(\bfQ+\mu^*_2\bfR\big)
		=\sup_{\mu_2\in\br}\big[-\mu_2+\lambda_{\min}\big(\bfQ+\mu_2\bfR\big],
	\end{equation}
	and $\bfx^*$ is a solution of the following system
	\begin{align}\label{eq:first-order}
		\begin{cases}
			\bfH(\mu_1^*,\mu_2^*)\bfx=\bf0\\
			\bfx^\top\bfx=1\\
			\bfx^\top\bfR\bfx=1.
		\end{cases}
	\end{align}
\end{thm}
\prf
For any $\mu_1, \mu_2\in\br$, the Courant-Fischer Theorem yields that
\begin{align}\label{eq:tmp}
	\begin{split}
		\lambda_{\min}\big(\bfH(\mu_1,\mu_2)\big)&=\min_{\|\bfx\|_2=1}\bfx^\top\bfH(\mu_1,\mu_2)\bfx\\
		&=\mu_1+\lambda_{\min}(\bfQ+\mu_2\bfR),
	\end{split}
\end{align}
which implies that
\[
\bfH(\mu_1,\mu_2)\succeq 0\iff\mu_1\ge -\lambda_{\min}(\bfQ+\mu_2\bfR).
\]
In particular, $\forall\mu_2\in\br$ and $\mu_1':=-\lambda_{\min}(\bfQ+\mu_2\bfR)$,
there holds $\bfH(\mu_1',\mu_2)\succeq 0$. Thus
\begin{align}\label{eq:lem:my1-tmp1}
		\eta^*&=\sup_{\bfH(\mu_1,\mu_2)\succeq 0}(-\mu_1-\mu_2)\nonumber\\
		&=\sup_{\mu_1\ge\mu_1',\mu_2\in\br}(-\mu_1-\mu_2)=\sup_{\mu_2\in\br}(-\mu_1'-\mu_2)\nonumber\\
		&=\sup_{\mu_2\in\br}\big[-\mu_2+\lambda_{\min}(\bfQ+\mu_2\bfR)\big].
\end{align}
\par
\textbf{Necessity:}~
If $\bfx^*$ is a solution of geQECQP, then by 
Lemma \ref{lemma}, there exists $\mu_1^*, \mu_2^*\in\br$ satisfying \eqref{eq:iffcondition},
which implies that 
$\lambda_{\min}\big(\bfH(\mu^*_1,\mu^*_2)\big)=0$
and
\[
\bfL(\bfx^*,\mu^*_1,\mu^*_2)=
\bfx^{*\top}\bfH(\mu^*_1,\mu^*_2)\bfx^*-\mu^*_1-\mu^*_2
=g(\mu^*_1,\mu^*_2).
\]
The first equality together with \eqref{eq:tmp} yields
$\mu^*_1=-\lambda_{\min}(\bfQ+\mu^*_2\bfR)$. 
The second equality and the weak duality of geQECQP yields that
\begin{align*}
	\zeta^*&:=\inf_{\bfx\in\mathcal{D}}\bfx^\top\bfQ\bfx
	\le\bfx^{*\top}\bfQ\bfx^*
	=\bfL(\bfx^*,\mu^*_1,\mu^*_2)\\
	&=g(\mu^*_1,\mu^*_2)\le\eta^*\le\zeta^*.
\end{align*}
Therefore
\[
\eta^*=g(\mu^*_1,\mu^*_2)=-\mu_1^*-\mu_2^*
=-\mu^*_2+\lambda_{\min}\big(\bfQ+\mu^*_2\bfR),
\]
which together with \eqref{eq:lem:my1-tmp1} deduces \eqref{eq:lem:my1-1} immediately.
\par
It is easy to see that $\bfx^*$ is a solution of the system \eqref{eq:first-order}.
\par
\textbf{Sufficiency:}~
Let $\mu_2^*\in\br$ and $\mu_1^*=-\lambda_{\min}(\bfQ+\mu^*_2\bfR)$
satisfy \eqref{eq:lem:my1-1} and $\bfx^*$ be a solution of \eqref{eq:first-order}.
Then by \eqref{eq:tmp} we have that $\lambda_{\min}\big(\bfH(\mu^*_1,\mu^*_2)\big)=0$,
which implies that $\bfH(\mu^*_1,\mu^*_2)\succeq 0$. Since
$\bfx^*$ is a solution of \eqref{eq:first-order}, we have that
$\bfx^*\in\mathcal{D}$ and $\bfH(\mu^*_1,\mu^*_2)\bfx^*=0$. 
By Lemma \ref{lemma}, $\bfx^*$ is the optimal solution of geQECQP. Proof completed. \bbox
\par
Theorem \ref{lem:my1} provides a way to solve the geQECQP \eqref{opt:gene_QECQP}:
\begin{description}
	\item [Step 1]: 
	Solve the optimization problem \eqref{eq:lem:my1-1}
	to find $\mu_2^*$ and let $\mu_1^*:=-\lambda_{\min}(\bfQ+\mu^*_2\bfR)$. 
	\item [Step 2]: Solve the system \eqref{eq:first-order} to find $\bfx^*$.
\end{description}
\par
According to
\[
\lambda_{\min}(\bfQ+\mu_2\bfR)
=\min_{\|\bfx\|_2=1}\bfx^\top(\bfQ+\mu_2\bfR)\bfx,
\]
we know that $f(\mu_2):=-\mu_2+\lambda_{\min}(\bfQ+\mu_2\bfR)$
is concave on $\br$. It shows that problem \eqref{eq:lem:my1-1} is a concave problem,
which can be solved by a sequence of bisection searches with the initial search interval determined by Weyl's inequality or empirically. 

\par
Let us come back to the QECQPs \eqref{opt:QECQP} for computing the grpah Fourier basis when $\bfA_k^\top\bfJ\bfA_k$ has both eigenvalues $1$ and $-1$. Let 
\begin{align*}
	\begin{cases}
		\bfQ:=\bfA_k^\top\bfL\bfA_k\in\br^{l\times l},\\
		\bfR:=\bfA_k^\top\bfJ\bfA_k+\bfI_l\in\br^{l\times l}.
	\end{cases}
\end{align*}
Then the eigenvalues of $\bfR$ are $\{0, 2\}$. This fact guarantees that the assumptions (a)--(c) of Lemma \ref{lemma} hold.

The pseudo-codes for \textbf{SolveOpt} that solve the QECQPs \eqref{opt:QECQP}
are presented in Algorithm \ref{alg:solveopt}, where the $3-4$ steps attempt to solve the system \eqref{eq:first-order}. The output $\xi^*$ is the global optimal solution of the problem \eqref{opt:QECQP} in each iteration.
\begin{algorithm}[htbp]
	\renewcommand{\algorithmicrequire}{\textbf{Input:}}
	\renewcommand{\algorithmicensure}{\textbf{Output:}}
	\caption{SolveOpt}
	\label{alg:solveopt}
	\begin{algorithmic}[1]
		\REQUIRE $\bfL,\bfJ,\bfA_k$
		\STATE Set $\bfQ=\bfA_k^\top\bfL\bfA_k,\bfR=\bfA_k^\top\bfJ\bfA_k+\bfI_l$
		\STATE Solve the unconstrained optimization problem \eqref{eq:lem:my1-1} by bisection method to obtain $\mu_1^*,\mu_2^*$ within a pre-specified tolerance.
		\STATE Compute an orthonormal basis of the null space of $\bfH(\mu_1^*,\mu_2^*)$ to form matrix $\bfB_H$
		\STATE Solve the equation $\mathbf{b}^\top\mathbf{b}=\mathbf{b}^\top(\bfB_H^\top\bfR\bfB_H)\mathbf{b}=1$ 
		\STATE Let $\xi^*=\bfB_H\mathbf{b}$
		\ENSURE $\xi^*$
	\end{algorithmic}
\end{algorithm}


\subsection{Perfect Reconstruction Equation}	

Now we have provided methods to design the sampling matrix $\bfJ$ and a graph Fourier basis matrix $\bfU$ such that $\bfJ\bfU=\bfU\Phi$, where $\Phi$ is a permutation matrix. Then any filter can be constructed by $\bfU$ and a filter vector, e.g., $\bfF_h=\bfU\diag(\bfh)\bfU^\top$. Accordingly, we can simplify the perfect reconstruction condition \eqref{eq:PR_2} similar to that in the case of bipartite graphs, as described in the following theorem:

\begin{thm}\label{coro:pr}
Let $\bfU$ be a  graph Fourier basis matrix and $\bfJ$ be a sampling matrix satisfying
\begin{equation}\label{eq:co:cond1}
\bfJ\bfU=\bfU\Phi
\end{equation}
for a symmetric permutation matrix $\Phi$. Then the perfect reconstruction condition \eqref{eq:PR_2} holds if the filter vectors $\bfh_0,\bfh_1,\bfg_0,\bfg_1$ satisfy
\begin{align}\label{eq:co:cond2}
\begin{cases}
\bfg_0\odot\bfh_0+\bfg_1\odot\bfh_1=2\bf1,\\
(\Phi\bfg_0)\odot\bfh_0-(\Phi\bfg_1)\odot\bfh_1=\bf0,
\end{cases}
\end{align}
where $\odot$ stands for the Hadamard product.
\end{thm}
	\prf
	Using $\bfU^\top\bfJ\bfU=\Phi$, the perfect reconstruction condition \eqref{eq:PR_2} can be written as 
	\begin{align}\label{eq:PR_3}
		\begin{cases}
			\diag(\bfg_0)\diag(\bfh_0)+\diag(\bfg_1)\diag(\bfh_1)=2\bfI_N,\\
			\diag(\bfg_0)\Phi\diag(\bfh_0)-\diag(\bfg_1)\Phi\diag(\bfh_1)=\bf0.
		\end{cases}
	\end{align}
Premultiply $\Phi$ on both sides of the second equation, then by $\Phi\diag(\bfh)\Phi=\diag(\Phi\bfh)$, we have
\begin{align*}
	\begin{cases}
		\diag(\bfg_0)\diag(\bfh_0)+\diag(\bfg_1)\diag(\bfh_1)=2\bfI_N,\\
		\diag(\Phi\bfg_0)\diag(\bfh_0)-\diag(\Phi\bfg_1)\diag(\bfh_1)=\bf0,
	\end{cases}
\end{align*}
which is equivalent to \eqref{eq:co:cond2}. Proof completed. 
\bbox

\subsection{Constructing the Filter Vectors}
Given a sampling matrix $\bfJ$, we have developed a method to compute a graph Fourier basis $\bfU$ satisfying $\bfJ\bfU=\bfU\Phi$, where $\Phi$ is a permutation matrix. According to Theorem \ref{coro:pr}, the construction of perfect reconstruction two-channel filterbanks is
reduced to finding the filter vectors $\bfh_0,\bfh_1,\bfg_0,\bfg_1$ satisfying 
\eqref{eq:co:cond2}.
\par
Similar to the design in \cite{Narang2013Compact}, we consider an orthogonal construction: 
let $\bfg_1=\Phi\bfh_0,\bfh_1=\Phi\bfg_0$ and $\bfg_0=\bfh_0$. Then \eqref{eq:co:cond2} reduces to 
\[
(\bfI_N+\Phi)(\bfh_0\odot \bfh_0)=2\bf1.
\]
\par
Denote $\bfh:=\bfh_0\odot\bfh_0$, we want it to be a lowpass filter vector satisfying the above equation. For example, one can minimize the error $\|\bfh-\bfh^{des}\|_\infty$ (other vector norms are also optional) to approximate a desired filter response $\bfh^{des}$:
\begin{align}\label{opt:filter}
	\begin{cases}
		\min\limits_{\bfh\in\br^N}&\|\bfh-\bfh^{des}\|_\infty,\\
		\st &(\bfI_N+\Phi)\bfh=2\bf1,
	\end{cases}
\end{align} 
or consider the design of $\bfh$ as follows:
\begin{align}\label{eq:filtervector}
	\begin{cases}
		\bfh(i)=h^*(i,j),& \Phi(i,j)=1,i<j,\\
		\bfh(j)=2-h^*(i,j),&\Phi(i,j)=1,i<j,\\
		\bfh(i)=1,&\Phi(i,i)=1,
	\end{cases}
\end{align}
for $1\leq i,j\leq N$, where $h^*(i,j)$ is a function of $(i,j)$ that provides more flexibility for adjusting the filter response. It is easy to verify that $\bfh$ satisfies the constraint $(\bfI_N+\Phi)\bfh=2\bf1$.

\subsection{Multiresolution Analysis Using Two-Channel Filterbank}
Given a graph signal $\bff$, the analysis filterbank will decompose it into a shorter lowpass component as a coarse resolution version of $\bff$ and a shorter highpass component that describes the details of $\bff$, as shown in Figure \ref{fig:2-channel-banks}. Analogous to the traditional filterbanks for time-domain signals, further decomposition can be recursively performed on the lowpass component to yield coarser approximations of $\bff$, as well as finer details.

Unlike the traditional situation, a reduced underlying graph is required for the construction of the filterbank in the next level of decomposition. A popular option is to use the Kron-reduction scheme \cite{Dorfler2012Kron} to reduce the graph. Start with the graph $\mcg=\{\mcv,\mce,\bfW\}$
with Laplacian matrix $\bfL$. If $\mcv_L\subset\mcv$ is the subset of vertices that we keep in the lowpass channel after downsampling, then the kron reduction of $\bfL$ related to $\mcv_L$ is given by
\begin{equation*}
	\bfL^{kr}:=\bfL_{\mcv_L,\mcv_L}-\bfL_{\mcv_L,\mcv_L^c}\bfL_{\mcv_L^c,\mcv_L^c}^{-1}\bfL_{\mcv_L^c,\mcv_L},
\end{equation*}
where $\mcv_L^c:=\mcv\backslash\mcv_L$ and $\bfL_{A,B}$ is the submatrix consisting of all the entries of $\bfL$ whose row index is in $A$ and whose column index is in $B$. Let $\bfW^{kr}:=\diag(\diag(\bfL^{kr}))-\bfL^{kr}$, then we can obtain a reduced graph $\mcg^{kr}:=\{\mcv_L,\mce^{kr},\bfW^{kr}\}$. The edge set $\mce^{kr}$ is derived from $\bfW^{kr}$.
Since repeated Kron reduction makes graphs denser and denser \cite{Dorfler2012Kron,Shuman2016Amultiscale}, we use a spectral sparsification algorithm of Spielman and Srivastava \cite{spielman2011graph} to do an edge sparsification.  Then the filterbank in the next level is constructed based on this sparsified graph.

\section{Experiments}\label{sec:expe}

\subsection{Instances of the Proposed Graph Fourier Basis}

In this section, we show the graph Fourier basis computed by the proposed method  on both bipartite and non-bipartite graphs. 

\begin{figure}[hbtp]
 	\centering{
 		\includegraphics[width=0.45\textwidth]{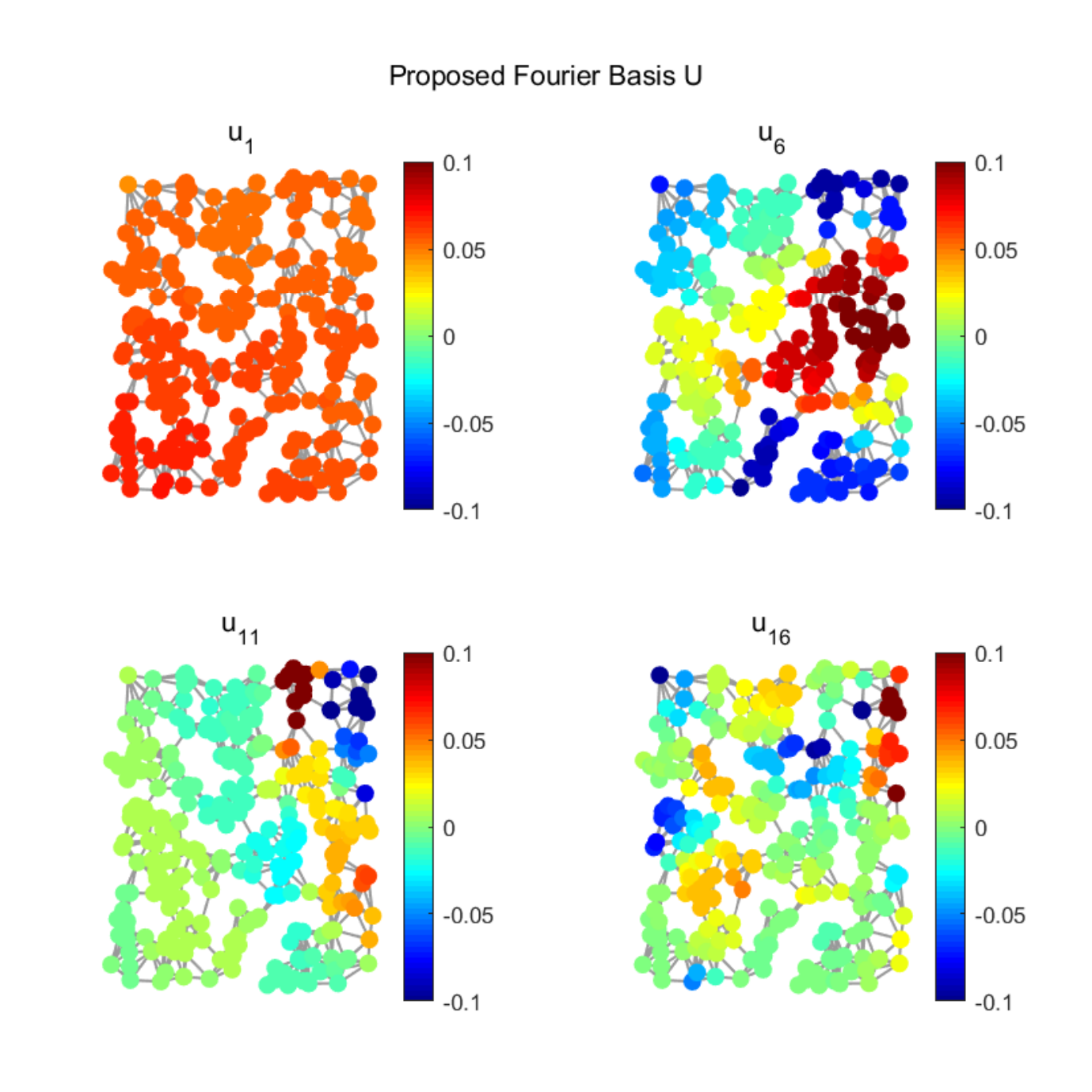}
 		\includegraphics[width=0.45\textwidth]{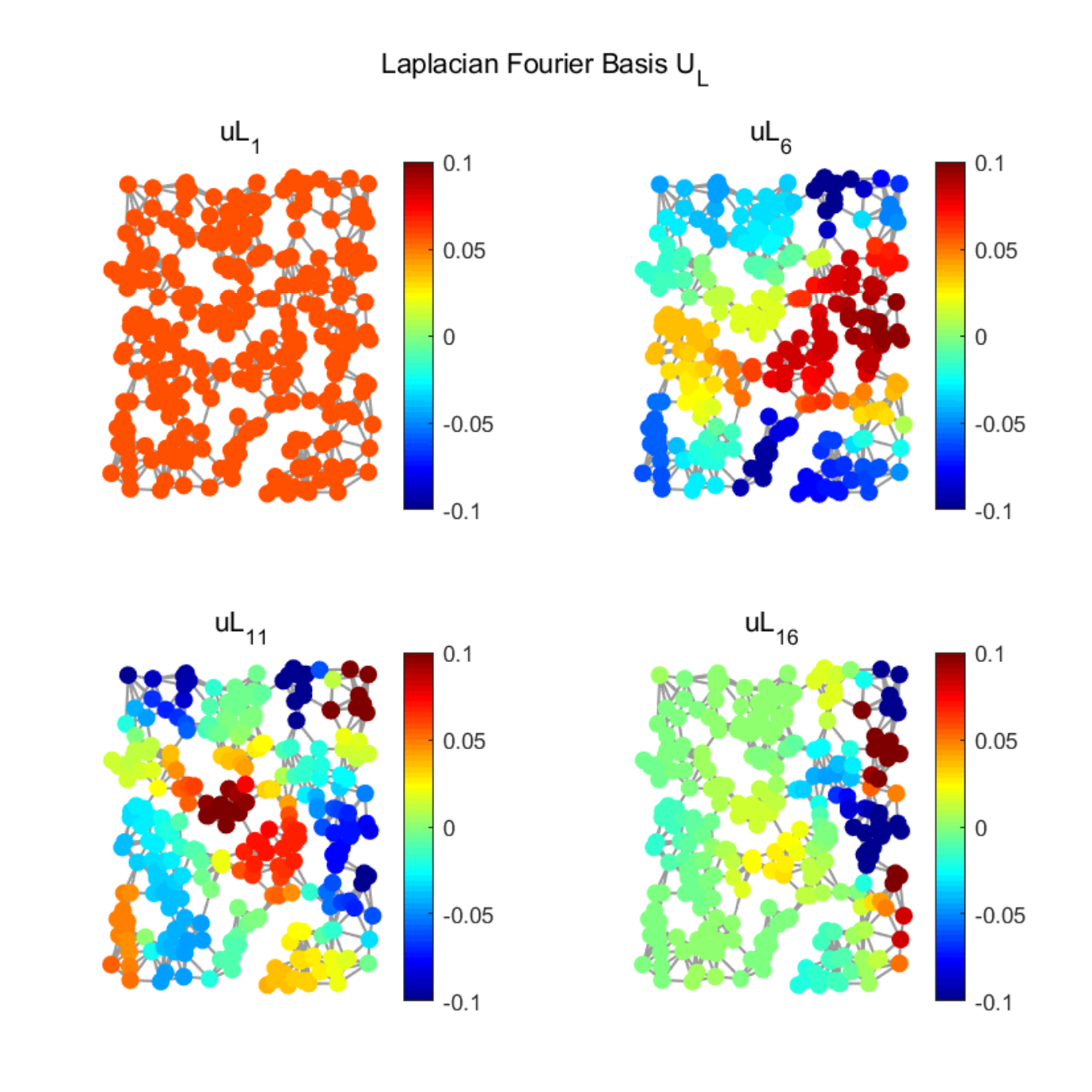}	
 		\caption{\small $\mcg$ is the sensor graph. Top Four: Some of the Fourier basis vectors computed by the proposed method. Bottom Four: Some of the eigenvectors of the Laplacian matrix.}\label{fig:UL}
 	}
 \end{figure}

Given a graph $\mcg$, the eigen-decomposition of its associated Laplacian matrix is given by $\bfL:=\bfU_L\Lambda_L{\bfU_L}^\top$, where $\Lambda_L:=\diag(\lambda_{L_1},\lambda_{L_2},\cdots,\lambda_{L_N})$ and the eigenvalues $\{\lambda_{L_i}\}_{i=1}^N$ are in ascending order. First of all, we need to construct a sampling matrix $\bfJ$, then feed $\bfJ,\bfL$ into Algorithm \ref{alg:basis} to compute a graph Fourier basis $\bfU$. If $\mcg$ is non-bipartite, we use Algorithm \ref{alg:J} to obtain $\bfJ$. Otherwise, suppose $\mcg$ is a bipartite graph with two parts $\mcv_1$ and $\mcv_2$, we downsample it by keeping one of $\mcv_1$ and $\mcv_2$ and discarding the other. In what follows, we denote $\bfu_i,\bfu_{L_i}$  the $i$-th column of $\bfU$ and $\bfU_L$, respectively.
\begin{figure}[hbtp]
	\centering{
		\includegraphics[width=0.24\textwidth]{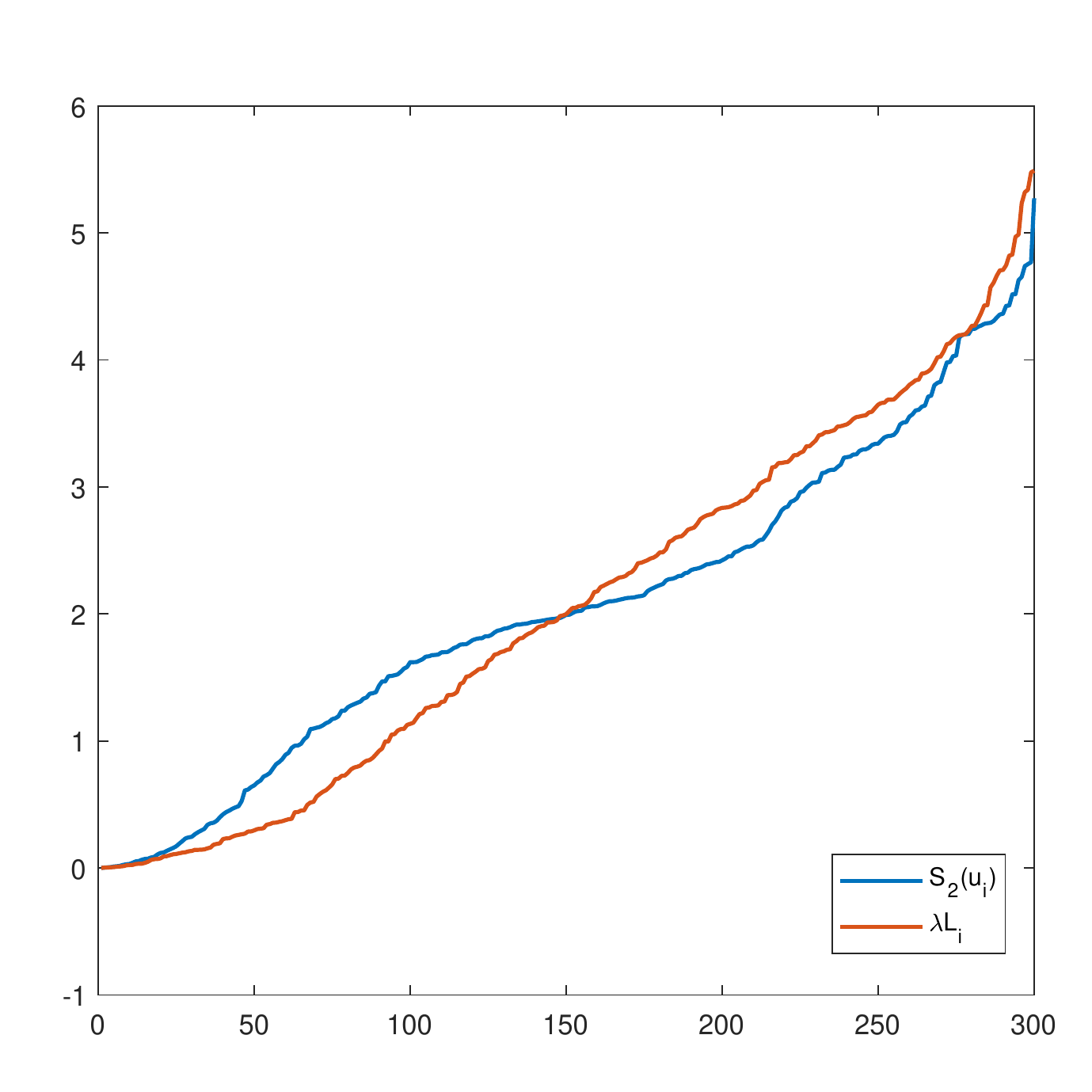}
		\includegraphics[width=0.24\textwidth]{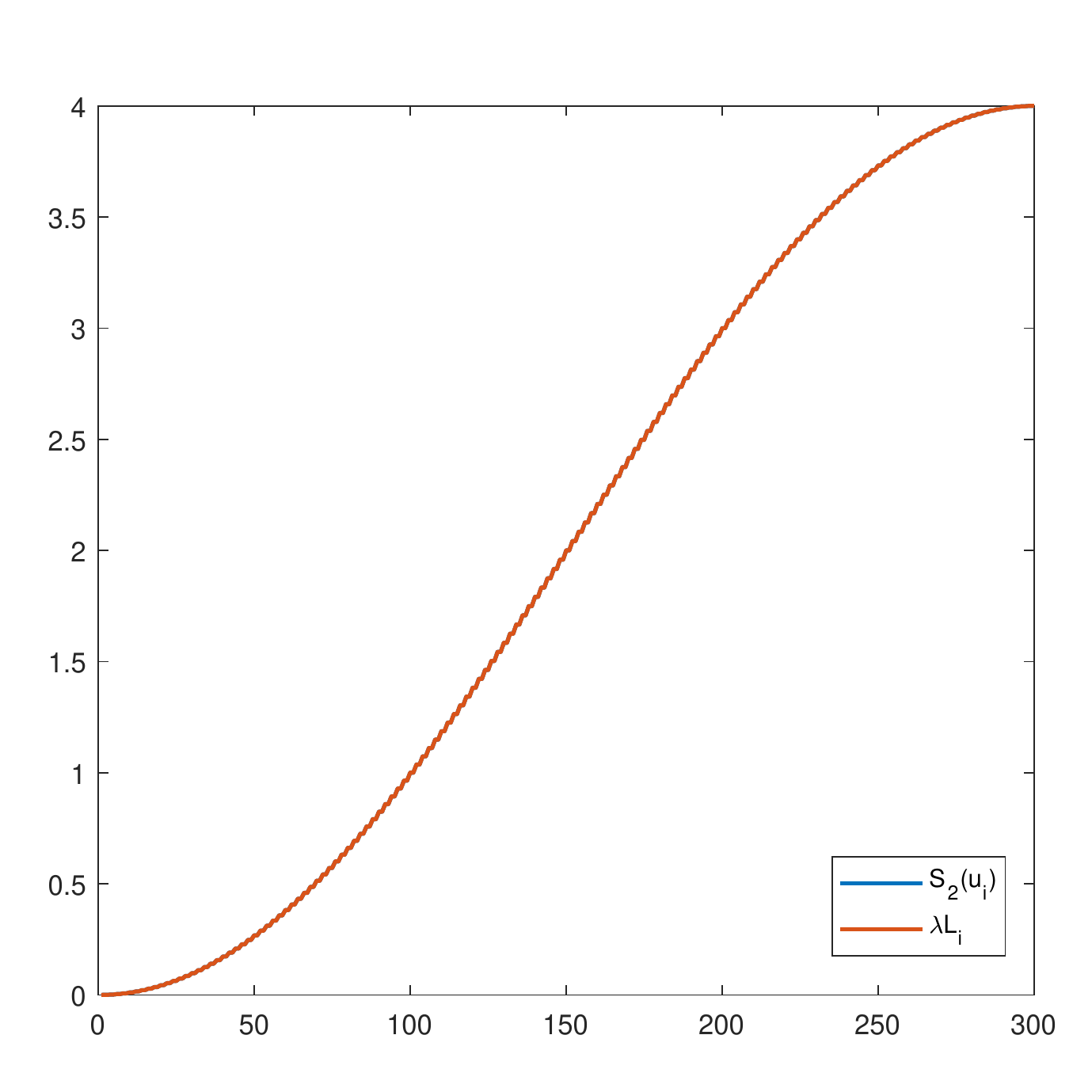}
		\caption{\small The oscillation of $\{\bfu\}_{i=1}^N$ and $\{\bfu_{L_i}\}_{i=1}^N$ respectively. Left: when $\mcg$ is the sensor graph; Right: when $\mcg$ is the ring graph.}\label{fig:eigenvals}
	}
\end{figure}
 First we consider the non-bipartite case. We use the GSP matlab toolbox \cite{perraudin2014gspbox} to generate a random sensor graph $\mcg$ with $300$ vertices.  Figure \ref{fig:UL} shows some of the columns of $\bfU$ and $\bfU_L$: $\{\bfu_{5i-4}\}_{i=1}^4$ and $\{\bfu_{L_{5i-4}}\}_{i=1}^4$. We can see that $\bfu_i$'s are different from $\bfu_{L_i}$'s and $\bfu_1$ is no longer a constant vector as $\bfu_{L_1}$ is, due to the constraint $\bfu_i^\top\bfJ\bfu_i=0$ or $\bfu_i^\top\bfJ\bfu_i=\pm1$. 
 
 \begin{figure}[hbtp]
 	\centering{
 		\includegraphics[width=0.45\textwidth]{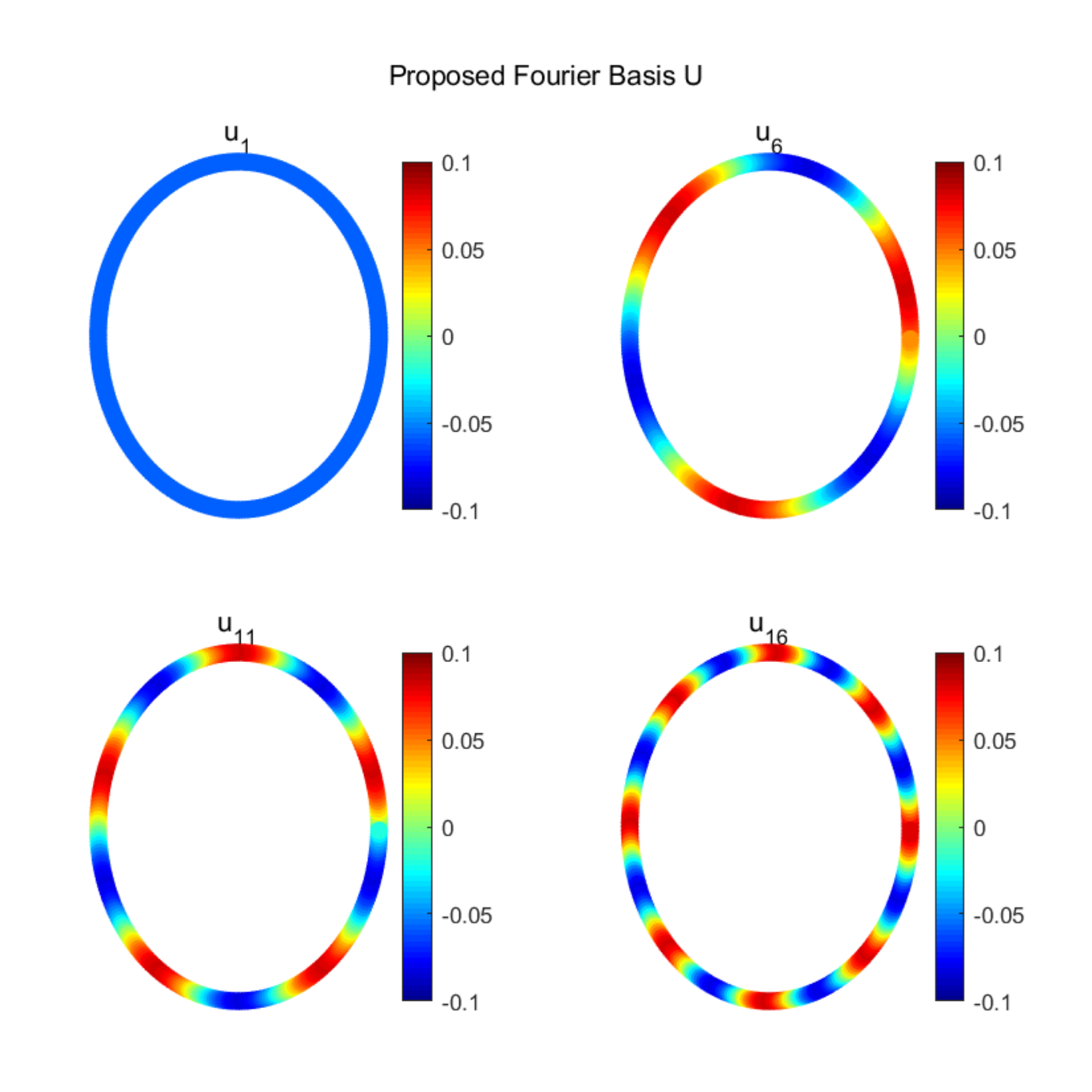}
 		\includegraphics[width=0.45\textwidth]{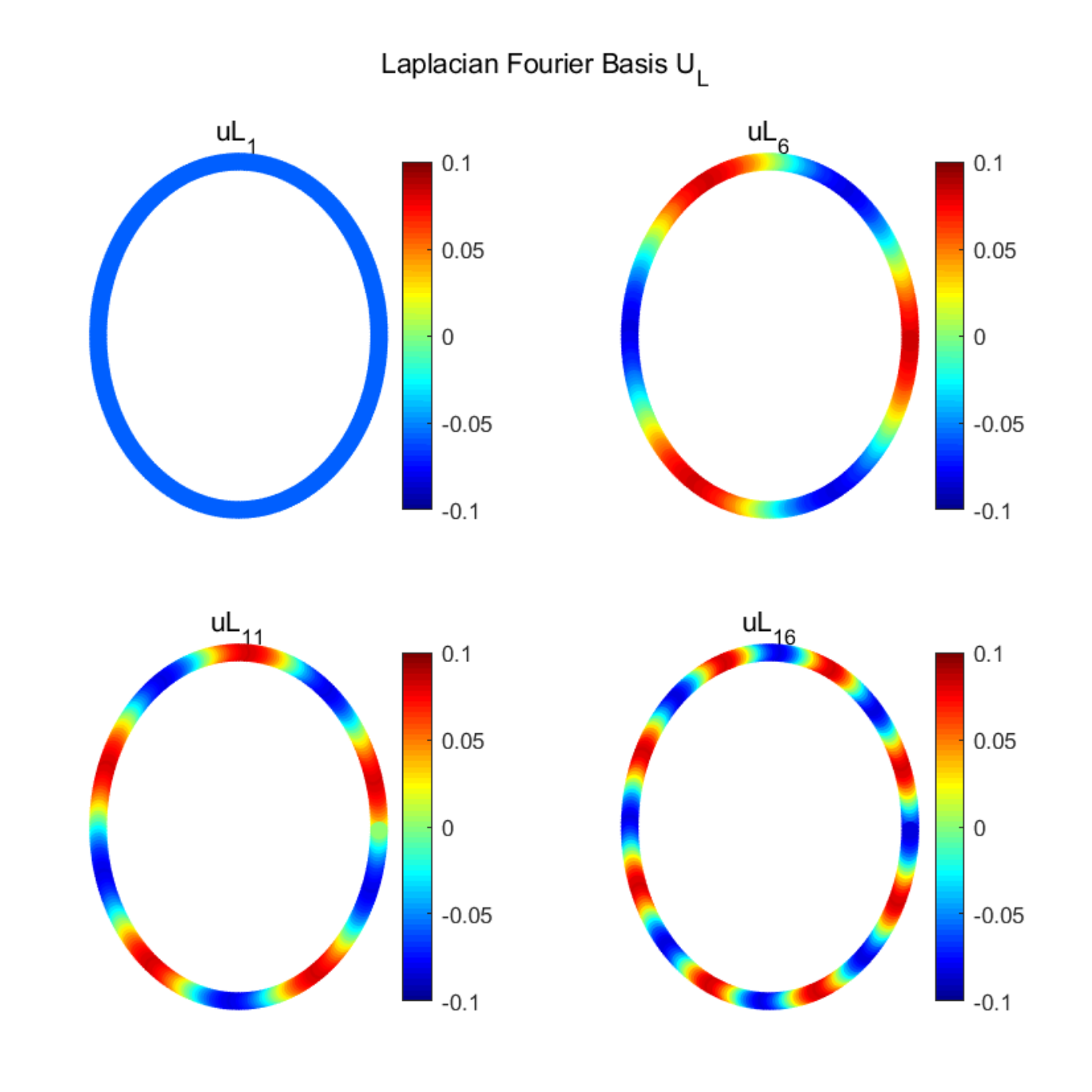}	
 		\caption{\small $\mcg$ is the ring graph. Top Four: Some of the Fourier basis vectors computed by the proposed method. Bottom Four: Some of the eigenvectors of the Laplacian matrix.}\label{fig:ringUL}
 	}
 \end{figure}
 
  The left image of Figure \ref{fig:eigenvals} shows how $\{\bfu\}_{i=1}^N$ and $\{\bfu_{L_i}\}_{i=1}^N$ oscillate w.r.t  $\mcg$ as $i$ increases. Specifically, the blue curve depicts the change of $\{S_2(\bfu_i)\}_{i=1}^N$ and the red curve depicts the change of $\{S_2(\bfu_{L_i})\}_{i=1}^N$. Note that $S_2(\bfu_{L_i})=\lambda_{L_i}$.  
  One may find that the middle section of the blue curve is relatively flat, it is because each time we execute \textbf{SolveOpt} to produce $\bfu_i$ and output both $\bfu_i$ and $\bfJ\bfu_i$ as part of the graph Fourier basis. Although $\bfu_i$ is the current smoothest unit vector that satisfies $\bfu_i^\top\bfJ\bfu_i=0$, $\bfJ\bfu_i$ may not be the least smooth vector currently,
  this leads to $S_2(\bfu_i)\geq S_2(\bfu_{L_i})$ in the low-frequency area, while $S_2(\bfu_i)\leq S_2(\bfu_{L_i})$ in the high-frequency area. It is a shortcoming, because the Fourier coefficients of a smooth signal, i,e, $\hat{\bff}:=\bfU^\top\bff$, may decay more slowly than 
$\hat{\bff}_L:=\bfU_L^\top\bff$. We hope to find ways to mitigate this problem in the future.

 For the bipartite case, we generate a ring graph $\mcg$ with $300$ vertices. We alternately downsample the vertices on $\mcg$, then any two vertices in the retained subset $\mcv_1\subset\mcv$ are not connected, and so is $\mcv^c:=\mcv\backslash\mcv_1$. Define the sampling matrix $\bfJ$ as follows:
 \begin{align*}
 J_{ii}=\begin{cases}
 	1,&i\in\mcv_1\\
 	-1,&i\in\mcv_1^c
 \end{cases}.
 \end{align*}
 Under this downsampling pattern, it can be expected that $\bfU_L$ will be a solution of our proposed method due to the spectral folding phenomenon of bipartite graphs \cite{chung1997spectral,narang2012perfect}. 
 To verify it, we still compute a graph Fourier basis $\bfU$ using Algorithm \ref{alg:basis} and compare it with $\bfU_L$. The right image of Figure \ref{fig:eigenvals} shows that $\{S_2(\bfu_i)\}_{i=1}^N$ and $\{S_2(\bfu_{L_i})\}_{i=1}^N$ are exactly the same. Figure \ref{fig:ringUL} shows part of the basis vectors: $\{\bfu_{5i-4}\}_{i=1}^4$ and $\{\bfu_{L_{5i-4}}\}_{i=1}^4$.

\subsection{Experiments on the Graph Signal Processing}
We design experiments to test the performance of the proposed filterbank on signal compression and denoising. We also compare it with the most related work---graphQMF and graphBior proposed by Narang and Ortega \cite{narang2012perfect, Narang2013Compact}.

\subsubsection{Signal Compression on Real-world Data Compared with Related Work}
In this section, we compare the performance of our proposed method, referred to as optFB, with graphQMF and graphBior on real-world data. 

We use the dataset FIRSTMM offered by Neumann etc. \cite{FIRSTMMDB}. FIRSTMM contains a number of graphs, each of which is a 3D model of an object. We select the $38$-th graph with $919$ vertices and denote it by $\mcg$. Node attributes are considered as the graph signal $\bff$ residing on $\mcg$. We compute the weight $w_{ij}$ of the edge $(v_i,v_j)$ as the inverse of the sum of edge attributes, which are the distance between vertices and the change of curvature.
\par
One-layer decomposition is performed on $\bff$ using optFB, graphQMF and graphBior, respectively, where the filter vectors are constructed according to \eqref{eq:filtervector} with $h^*(i,j)\equiv2$. For optFB, the sampling matrix $\bfJ$ is computed by Algorithm \ref{alg:J}. It preserves $439$ vertices.

 Codes for graphQMF and graphBior are available at \cite{codesForGRAPHQMFandGRAPHBIOR}, offered by Narang and Ortega. We equip graphQMF with the ideal kernel. For graphBior, let the lowpass filter kernel have $7$ zeros at $0$ and the highpass kernel have $9$ zeros at $2$. Both of these two methods use the algorithm called ``degree of saturation heuristics'' to compute the coloring of $\mcg$ and use Harary's decomposition to decompose $\mcg$ into a collection of bipartite subgraphs. In this experiment, $2$ subgraphs are produced, so the multi-dimensional separable wavelet filter bank (the definition is referred to \cite{narang2012perfect}) has $4$ channels: LL, LH, HL, HH channel, with $316, 218,287,98$ vertices retained, respectively.
 
\begin{figure}[hbtp]
  	\centering{
  		\includegraphics[width=0.45\textwidth]{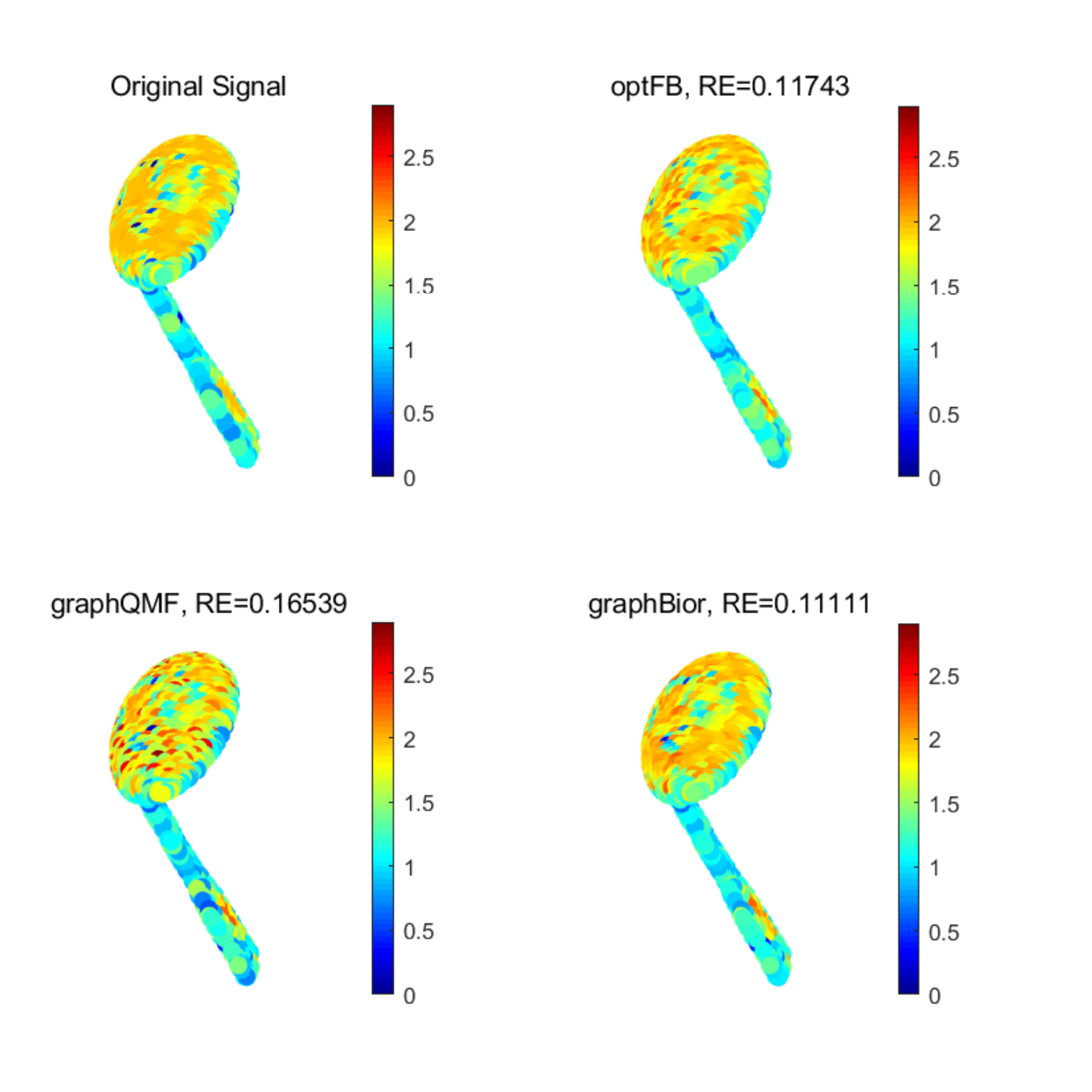}
  		\caption{\small Reconstructed graph signals from the lowpass channel of different methods. Top left: original signal; Top right: reconstruction from lowpass channel using optFB; Bottom left: reconstruction from LL and LH channels using graphQMF with an ideal kernel; Bottom right: reconstruction from LL and LH channels using graphBior with $7$ zeros at $0$ for the lowpass filter and $9$ zeros at $2$ for the highpass filter.}\label{fig:realdata}
  	}
  \end{figure}

 For optFB, we use the downsampled output of the lowpass channel containing $439$ samples for reconstruction. Whereas for graphQMF and graphBior, the downsampled outputs of LL and LH channels are used for reconstruction ($534$ samples in total). Figure \ref{fig:realdata} shows the relative errors $\mbox{RE}:=\frac{\|\bff-\hat{\bff}\|_2}{\|\bff\|_2}$ of each reconstructed signal $\hat{\bff}$. We can see that optFB and graphBior have similar relative errors and are smaller than that of graphQMF, but optFB is more efficient as it uses the least information for reconstruction. In terms of local analysis, graphBior has the best performance because its filters are polynomials in the normalized graph Laplacian. The local analysis capability of optFB remains to be studied and improved in our future work.
\par
We did not perform more than one layer of decomposition because in the case of graphQMF and graphBior, the multi-dimensional filterbank has too many channels, it is difficult to determine which bipartite subgraph to base on to design the sampling matrix and which channels to decompose next.

\subsubsection{Graph Signal Denoising on Arbitrary Graphs}
\begin{figure}[hbtp]
	\centering{
		\includegraphics[width=0.5\textwidth]{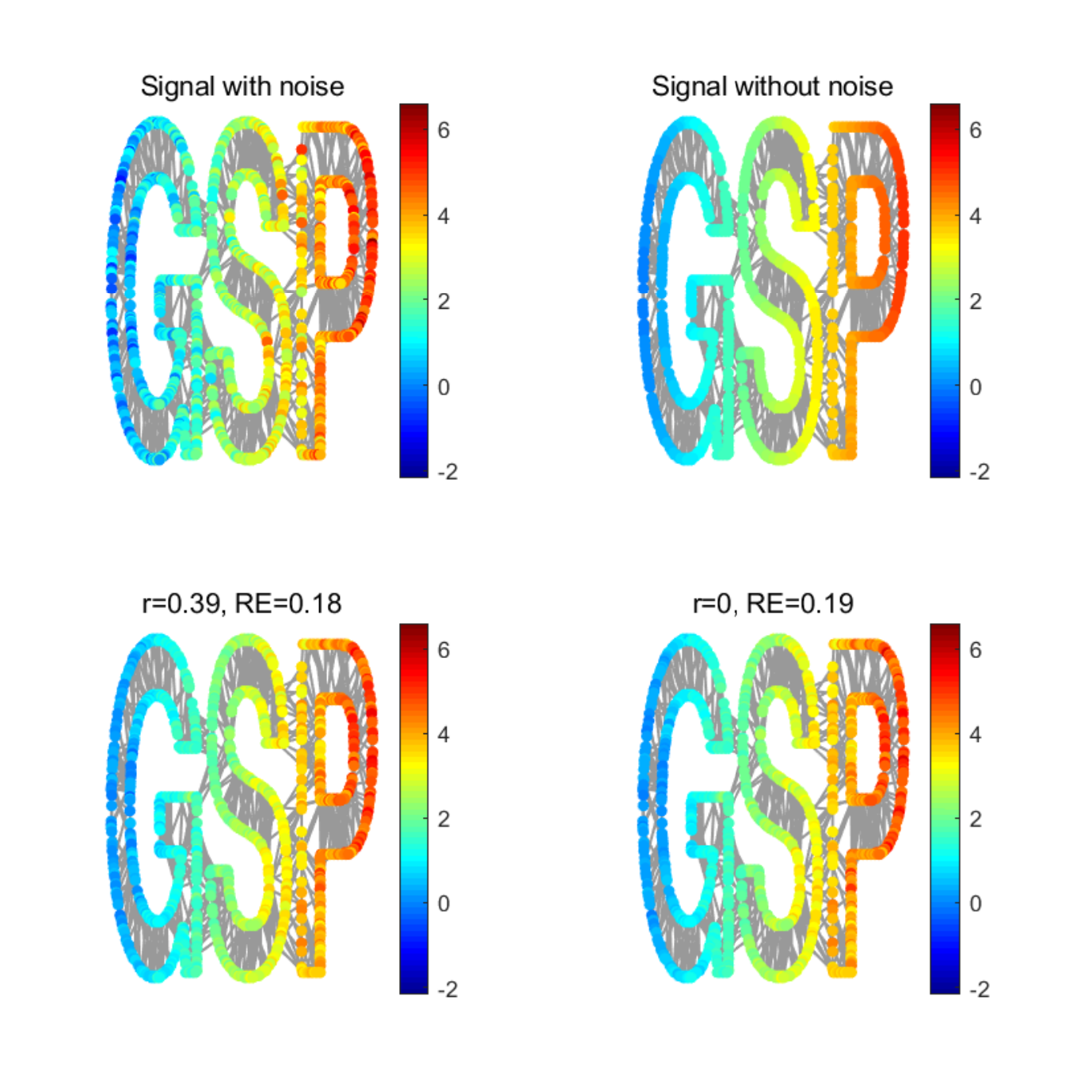}
		\caption{\small Reconstructions under different choices of $r$. Bottom left: $r$ is chosen to be the median of the absolute value of $\mathbf{e}$; Bottom right: $r=0$.}\label{fig:logoDenoise}
	}
\end{figure}

Let us use GSP toolbox to generate the logo graph $\mcg$ with $N=1130$ vertices, each vertex has a $2$D coordinate. Let $\bfx_c\in\br^N$ be the vector of $x$ coordinates of all vertices, we synthesize a graph signal $\bff_0$ on $\mcg$ as 
\[
\bff_0(i):=\frac{5(\bfx_c(i)-\min(\bfx_c))}{(\max(\bfx_c)-\min(\bfx_c))},~~i=1,...,N.
\]
Each sample of $\bff_0$ is highly correlated with the position of the corresponding vertex, so $\bff_0$ is smooth w.r.t $\mcg$. Additionally, we add Gaussian noise on $\bff_0$ to get a noisy signal 
\[
\bff:=\bff_0+\mathbf{e},
\]
where $\mathbf{e}$ is an i.i.d random vector following $\mathcal{N}(0,0.5)$ distribution. 
We perform $3$ layers decomposition on $\bff$, the filter vectors are designed according to \eqref{eq:filtervector} with $h^*(i,j)\equiv0$. Let $\bff_l^{(i)}$ and $\bff_h^{(i)}$ represent the downsampled lowpass and highpass component in the $i$-th layer, respectively. To denoise
the signal we adjust the highpass components $\{\bff_h^{(i)}\}_{i=1}^3$ of each layer with a preset 
threshold $r$:
\begin{align*}
	\tilde{\bff_h^{(i)}}(j):=\begin{cases}
		0,&|\bff_h^{(i)}(j)|>r,\\
		\bff_h^{(i)}(j),&\mbox{otherwise}.
	\end{cases}~~j=1,...,N
\end{align*}
and reconstruct a signal $\bff_1$ from the lowpass component $\bff_l^{(3)}$ and all the modified highpass components $\{\tilde{\bff_h^{(i)}}\}_{i=1}^3$. Figure \ref{fig:logoDenoise} shows the reconstructions under different choices of $r$.

\subsubsection{Locality Analysis of the Proposed Method}
To test the local analysis capability of optFB, we apply it to discontinuous graph signals residing on non-bipartite and bipartite graphs, respectively.

\begin{figure}[hbtp]
	\centering{
		\includegraphics[width=0.5\textwidth]{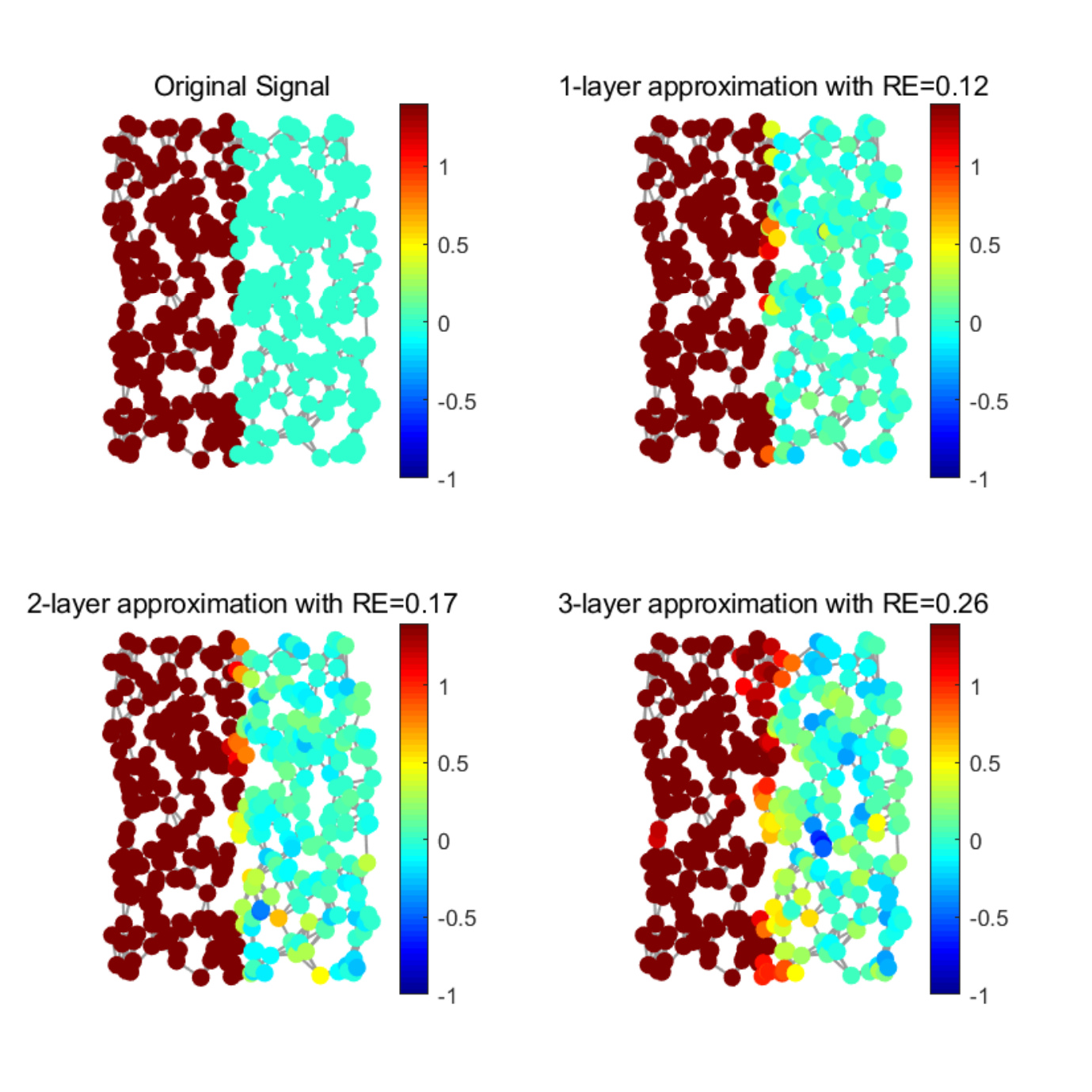}
		\caption{\small The ``$i$-th layer approximation'' represents the reconstructed signal from the downsampled lowpass component $\bff_l^{(i)}$ in the $i$-th layer decomposition. The underlying graph is $\mcg_s$.}\label{fig:local}
	}
\end{figure}

\begin{figure}[hbtp]
	\centering{
		\includegraphics[width=0.5\textwidth]{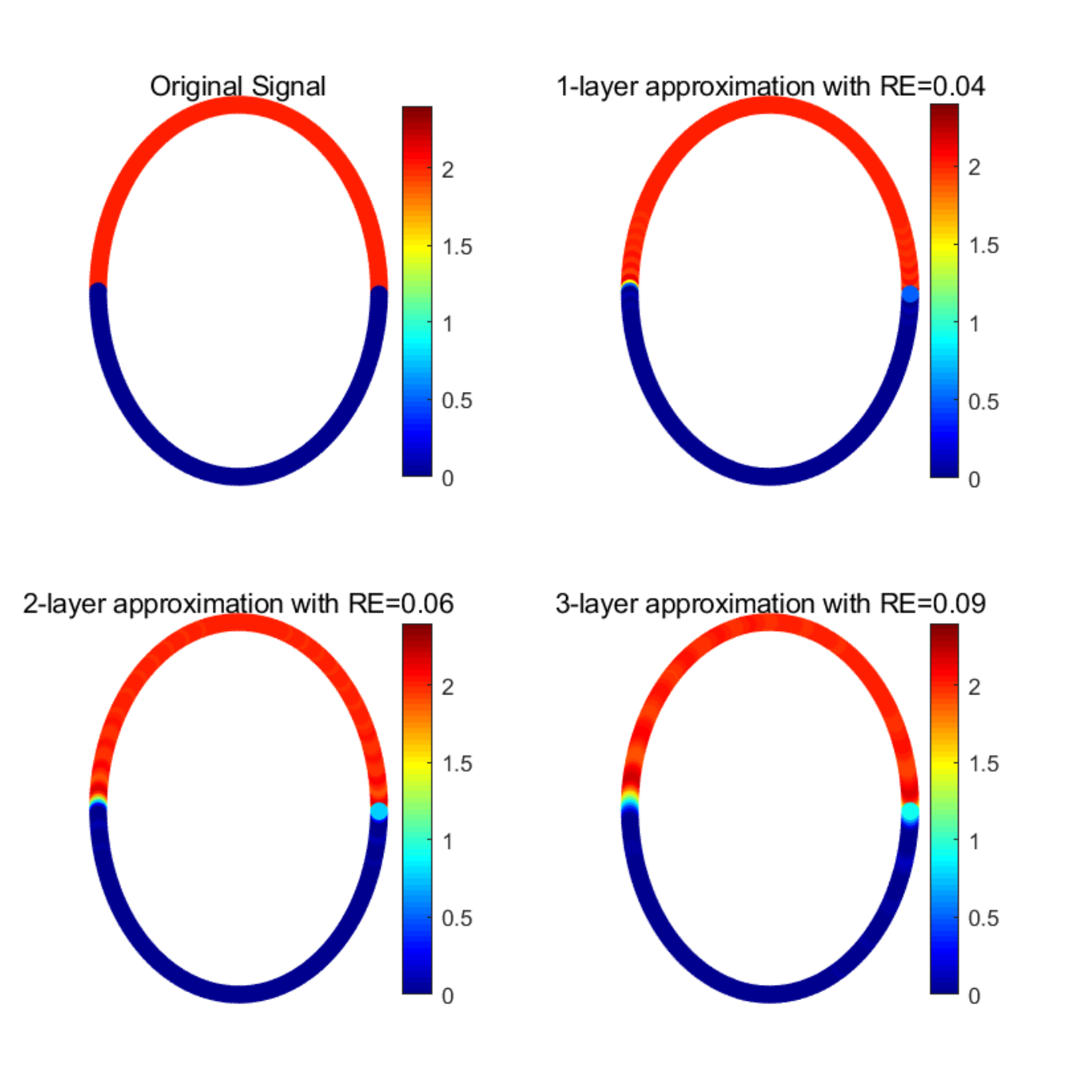}
		\caption{\small The $i$-th layer approximation represents the reconstructed signal from the downsampled lowpass component $\bff_l^{(i)}$ in the $i$-th layer decomposition. The underlying graph is $\mcg_r$.}\label{fig:ring_local}
	}
\end{figure}

Use GSP toolbox to generate a random sensor graph $\mcg_s$ and a ring graph $\mcg_r$, both of them have $N=400$ vertices. Generate signal $\bff$ as
\[
\bff(i)=\begin{cases}
	0,&i\leq N/2,\\
	2,&\mbox{otherwise}.
\end{cases}
\]
The filter vectors are constructed the same as before. Denote the downsampled lowpass component in the $i$-th layer decomposition as $\bff_l^{(i)}$. We show the reconstructions from $\bff_l^{(i)}, i=1,2,3$ in Figure \ref{fig:local} and Figure \ref{fig:ring_local}. If the filterbank is well localized in the vertex domain, the oscillation around the discontinuities will not spread too far. However, since the filters we use are not polynomials in $\bfL$, they may not have good locality in the vertex domain. As shown in the figures, the oscillation around the discontinuities of the reconstructed signals has large spread as $i$ increases.
How to design filters well localized in the vertex domain is the concern of our future work.

\section{Conclusion}
In this paper, we proposed the construction of perfect reconstruction two-channel filterbanks on arbitrary graphs without the operation of decomposing the graph into a collection of bipartite graphs. In order to achieve this, we compute a new graph Fourier basis through a series of quadratic equality constrained quadratic optimization problems, instead of using the eigenvectors of the graph Laplacian matrix. We gave an algorithm to compute the global optimal solution of these optimization problems. For the new graph Fourier basis, we proved a ``spectral folding phenomenon'' similar to that of bipartite graphs and constructed two-channel filterbanks based on it. However, the filters are not well localized in the vertex domain, thus how to improve the local analysis capability of the filters is the main concern of our future work. In addition, we also want to improve the proposed Fourier basis so that the smoothness of the basis vectors can have greater variance.

\newpage
\bibliographystyle{plain}
\bibliography{ref,UTF8-refs}

\end{document}